\documentclass[aps,prb,notitlepage,superscriptaddress,%
longbibliography,twocolumn]{revtex4-2}
\usepackage{natbib}
\usepackage{graphicx}
\usepackage{amssymb}
\usepackage{amsmath}
\usepackage{ifthen}
\usepackage{braket}
\usepackage{xcolor}
\usepackage{bm}
\usepackage{bbm}
\usepackage{hyperref}
\hypersetup{
colorlinks = true,
 urlcolor = blue,
 linkcolor = blue,
 citecolor = blue}
\usepackage{comment}
\usepackage{siunitx}
\usepackage{float}
\usepackage{dcolumn}
\newcolumntype{d}[1]{D{.}{.}{#1}}
\usepackage{subfigure}

\newcommand{\dd}{\mathrm{d}}
\newcommand{\ii}{\mathrm{i}}
\newcommand{\ee}{\mathrm{e}}

\renewcommand{\Re}{{\rm{Re}}}
\renewcommand{\Im}{{\rm{Im}}}

\newcommand{\au}{\mathrm{a.u.}}

\newcommand{\LD}{\mathrm{LD}}
\newcommand{\CM}{\mathrm{CM}}

\newcommand{\stdrule}{\rule[-2mm]{0mm}{6mm}}

\begin{document}

\title{Temperature-dependent dielectric function of 
intrinsic silicon: \\
Analytic models and atom-surface potentials}

\author{C. Moore}
\email{Email: 
christopher.moore@topologicalphysics.org}
\affiliation{Department of Physics and LAMOR, Missouri University of Science and
Technology, Rolla, Missouri 65409, USA}

\author{C. M. Adhikari}
\affiliation{Department of Chemistry, Physics and Materials Science,
Fayetteville State University, Fayetteville, North Carolina 28301, USA}

\author{T. Das}
\affiliation{Department of Physics and LAMOR, Missouri University of Science and
Technology, Rolla, Missouri 65409, USA}

\author{L. Resch}
\affiliation{Department of Physics and LAMOR, Missouri University of Science and
Technology, Rolla, Missouri 65409, USA}

\author{C. A. Ullrich}
\affiliation{Department of Physics and Astronomy, 
University of Missouri, Columbia, Missouri 65211, USA}

\author{U. D. Jentschura}
\affiliation{Department of Physics and LAMOR, Missouri University of Science and
Technology, Rolla, Missouri 65409, USA}

\begin{abstract}
The optical properties of monocrystalline, intrinsic silicon are of interest for 
technological applications as well as fundamental studies of atom-surface
interactions. For an enhanced understanding, it is of great interest to 
explore analytic models 
which are able to fit the experimentally determined dielectric
function $\epsilon(T_\Delta, \omega)$, over a wide range of frequencies
and a wide range of the temperature parameter $T_\Delta = (T-T_0)/T_0$, 
where $T_0 = 293\,{\rm K}$ represents room temperature.
Here, we find that a convenient functional form for the 
fitting of the dielectric function of silicon 
involves a Lorentz-Dirac curve with a complex, frequency-dependent amplitude 
parameter, which describes radiation reaction.
We apply this functional form to the expression 
$[\epsilon(T_\Delta, \omega) -1]/[ \epsilon(T_\Delta, \omega)+2]$,
inspired by the Clausius-Mossotti relation.
With a very limited set of fitting parameters, we are able to 
represent, to excellent accuracy, experimental data in the 
(angular) frequency range $0 < \omega < 0.16 \, {\rm a.u.}$
and $0< T_\Delta < 2.83$, corresponding to the 
temperature range $ 293\,{\rm K} < T < 1123\, {\rm K}$.
Using our approach, we evaluate the 
short-range $C_3$ and the long-range $C_4$ coefficients for the
interaction of helium atoms with the silicon surface. 
In order to validate our results, we compare to
a separate temperature-dependent direct fit of  $\epsilon(T_\Delta, \omega)$
to the Lorentz-Dirac model.
\end{abstract}

\maketitle



\section{Introduction}
\label{sec1}

Because of its enormous technological importance, 
the optical properties of monocrystalline, undoped
silicon, sometimes referred to 
as {\em intrinsic silicon}, have been
investigated in great detail over the 
past decades~\cite{MFRo1955,%
MFEtAl1958,%
Hu1975,%
IcPlWo1976,%
BaPi1979,%
JeMo1983,%
AsSt1983,%
Ed1985,%
Gr1987tt,%
As1988,%
LaEtAl1987,%
ToAd1991,%
Je1992,%
VuEtAl1993,%
BuBrWa1994,%
GrKr1995,%
KeGr1995,%
HeEtAl1998,%
SiHoHu1998,%
Gr2008,%
DeJo2012,%
PrKa2014,%
ScEtAl2015,%
KrEtAl2016cqg,%
ChBaJu2019}.
The determination of an appropriate analytic 
model for the frequency-dependent, 
and temperature-dependent, dielectric function 
also is of prime interest, especially 
because it may give insight into the 
physical mechanism that generates 
the response of the medium~\cite{DeJo2012,PrKa2014}.
In general, it is of obvious interest to 
find a satisfactory representation 
of the available data for the dielectric function
of silicon, using the most simple analytic 
functional form possible.
{\color{black} The aims of our paper are as follows: 
{\em (i)} We explore the applicability 
of simple functional forms, which we refer to as the 
Clausius-Mossotti and
Lorentz-Dirac models 
(which include radiation reaction damping terms) for the frequency-
and temperature-dependent
dielectric function of silicon.
{\em (ii)} We aim to describe the temperature
dependence of the dielectric function of intrinsic silicon, 
using an efficient model, i.e., using 
a small number of fitting parameters. Finally,
{\em (iii)} we aim to demonstrate the applicability of the 
functional forms of the temperature- and frequency-dependent 
dielectric function for the
calculation of a practically important quantity, namely, the 
short-range ($C_3$) and 
long-range ($C_4$) coefficients of the atom-surface
interaction for a few simple atomic systems interacting 
with intrinsic silicon.}

{\color{black} We have carefully examined available 
data sets for the real and imaginary 
parts of the dielectric function of silicon
and base our investigations on 
Refs.~[\onlinecite{Je1992,Gr2008,ScEtAl2015,VuEtAl1993,SiHoHu1998}]
(see also a pertinent comprehensive discussion in Appendix~\ref{appa}).
For these data sets, 
which cover the temperature range $ 293\,{\rm K} < T < 1123\, {\rm K}$,}
we attempt to find a uniform, 
simple, temperature-dependent analytic model 
for the dielectric function of monocrystalline
(intrinsic) silicon. Our motivation is
twofold. First, such an analytic model could
be of interest for practical applications,
and second, the most appropriate functional form 
for the description of the dielectric function
might otherwise give insight into the physical 
mechanism underlying the optical response of the 
medium. In Ref.~[\onlinecite{DeJo2012}],
it is pointed out that a two-resonance 
analytic model of the Lorentz-Dirac (LD) type 
can successfully describe the experimental 
data for the Si dielectric function 
over wide frequency ranges.
A physical interpretation 
and justification for the functional 
form used in Ref.~[\onlinecite{DeJo2012}]
is given in Refs.~[\onlinecite{PrKa2014,ChBaJu2019}].
This justification~\cite{DeJo2012,PrKa2014,ChBaJu2019} 
is based on the so-called Lorentz–Dirac force
(see Sec.~8.6.2 of Ref.~[\onlinecite{Je2017book}]
and Appendix~\ref{appb}). 
A second fitting method, which we also apply here,
tries to augment the Lorentz-Dirac approach 
using a functional form inspired by the 
Clausius-Mossotti (CM) relation.
The aim of the latter approach is to take into account the local-field 
effect inside the crystal.
The dual fitting method 
has been used in Ref.~[\onlinecite{OuCa2003}] 
where it has been shown that the Lorentz-Dirac and 
Clausius-Mossotti functional forms (without the radiation reaction 
term) can be mapped onto each other on the basis 
of a simple resonance frequency shift detailed in Eq.~(12) 
of Ref.~[\onlinecite{OuCa2003}].
(We note that the Clausius-Mossotti functional form
is referred to as the Lorentz-Lorenz formula 
in Ref.~[\onlinecite{OuCa2003}].) Here, we aim to explore 
if similar conclusions can be drawn when the model
is augmented by a radiation reaction term 
in the numerator of the resonance functional forms,
and in an application to the dielectric function
of a real material rather than a model problem.
A further motivation for our study comes from the 
fact that a number of density-functional theory (DFT) and
Bethe-Salpeter based 
approaches~\cite{OnReRu2002,BoEtAl2007rpp,Ul2012,Ul2015,MaReCe2016,BySuUl2020},
TDDFT \cite{BoEtAl2007rpp,Ul2012,Ul2015,BySuUl2020},
as well as QED-TDDFT frameworks~\cite{RuMaBa2011,To2013,FlRuApRu2017,FlEtAl2019}
suggest that the excitonic mechanism governing the 
dielectric function of silicon supports a functional form 
of the type explored here.

In order to ramify the motivation for
our investigations, let us mention two additional 
aspects originating in fundamental physics.
The first is the potential use of silicon
in gravitational wave detection
experiments, where an accurate understanding of the 
optical properties is crucial to gauge the 
achievable interferometric contrast~\cite{KrEtAl2016cqg}.
The second is the use of monocrystalline silicon 
as a substrate for atom-surface studies,
notably, at the Heidelberg Spin-Echo
Atomic Beam Apparatus (see Refs.~[\onlinecite{DKDuHaLa2001,DrDK2003,JeJaDK2016pra}]).
We here evaluate the 
temperature dependence of the Casimir coefficients $C_3$ (short-range) 
and $C_4$ (long-range), which represent 
the asymptotics of the atom-surface interaction energy,
for helium (and other) atoms interacting with monocrystalline
silicon. Helium has been 
of prime experimental interest and takes a very special
role in atom-surface studies~\cite{Ra1929,St1929,EsSt1930,FrSB2021},
and we devote special attention to the helium system here.
Details of other atoms are relegated to 
Ref.~[\onlinecite{MoEtAl2022suppl}].
{\color{black} We can anticipate that our two fitting methods lead to 
consistent numerical
results for the short-range $C_3$, and 
long-range $C_4$ coefficients.}

The paper is organized as follows.
In Sec.~\ref{sec2}, we discuss 
the fitting of the dielectric function
of intrinsic silicon to convenient
functional forms. Specifically, in Sec.~\ref{sec2A},
a functional form (a ``master function'') 
is indicated which will be used for our fitting 
in the following. 
(The physical justification of the ``master function''
is discussed on the basis of the Lorentz-Dirac equation.)
In Sec.~\ref{sec2B}, we discuss an approach to the 
fitting of the temperature-dependent  
dielectric function of silicon, which we refer 
to as the Clausius-Mossotti approach.
This approach is based on the comparison of a specific
ratio involving the dielectric function, 
to a generalized Lorentz-Dirac functional form,
with complex oscillator strengths.
The latter functional form constitutes
our ``master function''.
In Sec.~\ref{sec2C}, we discuss, for comparison,
an alternative approach to the description 
of the temperature-dependent dielectric function
based on the Lorentz-Dirac approach.
In Sec.~\ref{sec3}, we perform the 
evaluation of the coefficients $C_3$ and $C_4$
for helium interacting with silicon.
Conclusions are presented in Sec.~\ref{sec:conclu}.
SI mksA units are used throughout the paper.

\begin{table}[t!]
\caption{\label{table1}
Coefficients resulting from Clausius-Mossotti fitting,
as described in Eqs.~\eqref{eq:reFitCM} 
and~\eqref{eq:imFitCM},
are given for the first resonance of 
monocrystalline silicon
over a range of temperatures 
$0 < T_\Delta < 2.83$.
Here, $E_h$ is the Hartree energy and 
$\hbar$ is Planck's constant.}
\renewcommand{\arraystretch}{1.25}
\begin{center}
\begin{minipage}{1.00\linewidth}
\begin{tabular}
{c@{\hspace{0.06\linewidth}}
c@{\hspace{0.06\linewidth}}
c@{\hspace{0.06\linewidth}}
c@{\hspace{0.06\linewidth}}c}
\toprule\\[-1em]
$T_\Delta$ &
$a_1^\CM$ &
$\omega_1^\CM\,[E_h/\hbar]$ &
$\gamma_1^\CM\,[E_h/\hbar]$ &
$\gamma^{\prime\CM}_1\,[E_h/\hbar]$ \\[1ex]
\hline
0.000 & 0.004943 & 0.1293 & 0.01841 & 0.1306 \\
0.273 & 0.004856 & 0.1277 & 0.01973 & 0.1392 \\
0.444 & 0.004564 & 0.1266 & 0.01964 & 0.1474 \\
0.614 & 0.004715 & 0.1264 & 0.02030 & 0.1403 \\
0.785 & 0.004508 & 0.1258 & 0.02059 & 0.1440 \\
0.956 & 0.004647 & 0.1256 & 0.02075 & 0.1355 \\
1.126 & 0.004586 & 0.1249 & 0.02139 & 0.1405 \\
1.397 & 0.004903 & 0.1247 & 0.02221 & 0.1284 \\
1.468 & 0.005163 & 0.1243 & 0.02287 & 0.1173 \\
1.638 & 0.005588 & 0.1237 & 0.02529 & 0.1179 \\
2.321 & 0.007875 & 0.1242 & 0.03141 & 0.0614 \\
2.833 & 0.008155 & 0.1231 & 0.03376 & 0.0608 \\
\hline
\hline
\end{tabular}
\end{minipage}
\end{center}
\end{table}

%
%
\section{Dielectric Function of Silicon}
\label{sec2}

%
%
\subsection{Lorentz-Dirac and Master Function}
\label{sec2A}

In Refs.~[\onlinecite{DeJo2012,PrKa2014,ChBaJu2019}], the authors advocate 
to fit experimental data for the dielectric function of a 
reference material via a functional form of the Lorentz-–Dirac type, which is essentially equal to the Sellmeier form~\cite{Se1872}, but with a complex amplitude parameter (which could be understood as a complex oscillator strength), which takes the radiation reaction into account. {\color{black} Details of the derivation of the functional form have been 
discussed at length in the literature, and they are recalled 
for the convenience of the reader in Appendix~\ref{appb}
where we lay special emphasis on the sign of the imaginary 
part of the numerator term. As a result of these considerations,
we are motivated to define the functional form 
$f(T_\Delta, \omega)$,}
which we refer to as the Lorentz–Dirac master function, as follows:
\begin{equation}
\label{master}
f(T_\Delta, \omega) = \sum_{k=1}^{k_{\rm max}}
\frac{a_k(\omega_k^2 - \ii \gamma_k'\omega)}%
{\omega_k^2 - \omega^2 - i\omega \gamma_k} \,,
\end{equation}
with the dimensionless temperature parameter
\begin{equation}
T_\Delta = \frac{T - T_0}{T_0} \,,
\end{equation}
where $T_0 = 293$ K. In Eq.~\eqref{master}, the resonance energies $\omega_k$, 
the radiation reaction damping constants $\gamma_k'$ and level widths 
$\gamma_k$, and the amplitudes $a_k$ all depend on $T_\Delta$. The functional 
form given in Eq.~\eqref{master} has
a propagator denominator equal to that of a damped harmonic oscillator while
the numerator (the oscillator strength) has a nonvanishing imaginary part. The
parameter $k_{\rm max}$ terminates the sum over the generalized damped
oscillator terms; as we will show, the sum over oscillators leads to a
satisfactory representation of the dielectric function with only few terms,
resulting in $k_{\rm max}$ being a small integer.

\begin{table}[t!]
\caption{\label{table2}
We present the analog of Table~\ref{table1}
for the second resonance of 
monocrystalline silicon.}
\renewcommand{\arraystretch}{1.25}
\begin{center}
\begin{minipage}{1.00\linewidth}
\begin{tabular}
{c@{\hspace{0.06\linewidth}}
c@{\hspace{0.06\linewidth}}
c@{\hspace{0.06\linewidth}}
c@{\hspace{0.06\linewidth}}c}
\toprule\\[-1em]
$T_\Delta$ &
$a_2^\CM$ &
$\omega_2^\CM\,[E_h/\hbar]$ &
$\gamma_2^\CM\,[E_h/\hbar]$ &
$\gamma^{\prime\CM}_2\,[E_h/\hbar]$ \\[1ex]
\hline
0.000 & 0.7709 & 0.3117 & 0.0990 & 0.0971 \\
0.273 & 0.7739 & 0.3135 & 0.1066 & 0.1057 \\
0.444 & 0.7761 & 0.3135 & 0.1173 & 0.1176 \\
0.614 & 0.7766 & 0.3133 & 0.1191 & 0.1193 \\
0.785 & 0.7780 & 0.3138 & 0.1242 & 0.1247 \\
0.956 & 0.7783 & 0.3129 & 0.1293 & 0.1307 \\
1.126 & 0.7796 & 0.3136 & 0.1336 & 0.1351 \\
1.397 & 0.7804 & 0.3130 & 0.1363 & 0.1381 \\
1.468 & 0.7815 & 0.3122 & 0.1424 & 0.1447 \\
1.638 & 0.7847 & 0.3128 & 0.1333 & 0.1295 \\
2.321 & 0.7869 & 0.3072 & 0.1194 & 0.1121 \\
2.833 & 0.7949 & 0.3117 & 0.1159 & 0.1016 \\
\hline
\hline
\end{tabular}
\end{minipage}
\end{center}
\end{table}

\begin{figure*}[t]
\begin{center}
\includegraphics[width=1.98\columnwidth]{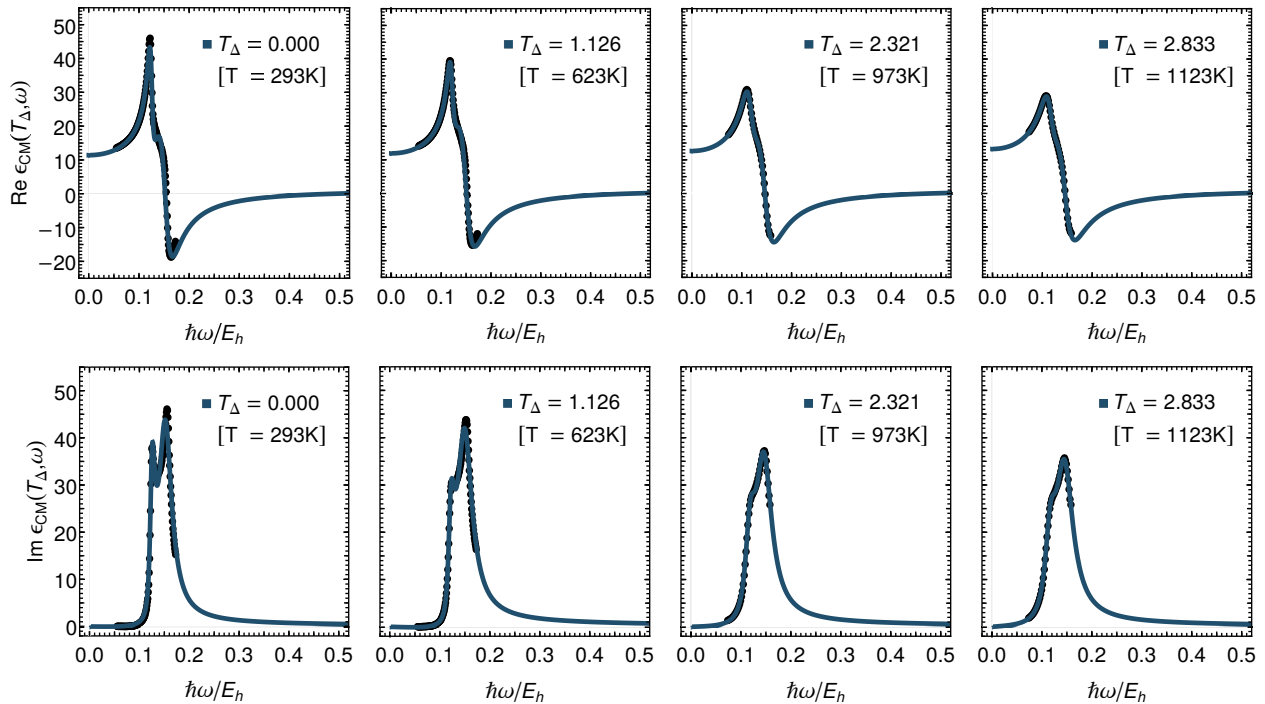}
\end{center}
\caption{\label{fig1} 
Real (top row) and Imaginary (bottom row) parts of the dielectric
function $\epsilon_\CM(T_\Delta, \omega)$ 
are plotted as functions of the angular frequency $\omega$, 
for monocrystalline (undoped, intrinsic) 
silicon for various temperatures. 
Experimental data 
\cite{ScEtAl2015,Je1992} (dotted) is shown to be in agreement with 
the CM fit (blue) defined in Eq.~\eqref{CMfit} 
for temperature-dependent parameters given in
Eq.~\eqref{eq:coefTempCM} and coefficients in Table~\ref{table3}. 
Note that the quantity $\hbar \omega/E_h$ is equal to the
angular frequency expressed in atomic units.  Eight additional temperatures, 
between $T_\Delta = 0$ and $T_\Delta = 2.83$, namely, the values
$T_\Delta = 0.273$, $0.444$, $0.614$, $0.785$, $0.956$, $1.297$, 
$1.468$, and $1.638$,
are considered in Ref.~\cite{MoEtAl2022suppl}.}
\end{figure*}

Before we discuss the actual fitting procedure, it is instructive to ask how
the specific form of $f(T_\Delta, \omega)$ can be justified from
first-principles theory. The {\em ab initio} calculation of the dielectric
function of a material is a two-step process: first, the electronic band
structure is obtained, either using density-functional theory (DFT) or Green's
function based approaches~\cite{OnReRu2002,MaReCe2016}. In the second step, the
band structure is taken as input to obtain the optical excitation spectrum of
the material via linear-response theory. To reproduce the double-peak structure
of the optical absorption spectrum of Si, it is essential to capture excitonic
effects. This can be accomplished using the Bethe-Salpeter equation
\cite{OnReRu2002,MaReCe2016} or time-dependent density-functional theory
(TDDFT) \cite{BoEtAl2007rpp,Ul2012,Ul2015,BySuUl2020}.  In both approaches,
one first constructs a noninteracting response function and then builds in
dynamical many-body effects, most notably the screened electron-hole
interactions.  The noninteracting response function features energy
denominators of exactly the same form as in Eq.~\eqref{master}. 
It is customary to choose
empirical line broadening parameters (corresponding to our $\gamma_k$) on the
order of 0.1 - 0.2 eV to obtain optical spectra in good agreement with
experiment.  This simulates the lifetime broadening caused by phonons,
disorder, or finite quasiparticle lifetimes \cite{CaEtAl2011prb}.

\begin{table}[t!]
\caption{\label{table3}
We present coefficients associated with 
the Clausius-Mossotti model given in
Eq.~(\ref{eq:coefTempCM}), found from fitting 
the coefficients in Table~\ref{table1} and~Table~\ref{table2}
as functions of temperature,
for monocrystalline silicon.}
\newcommand\colW{3em}
\catcode`"=9
\renewcommand{\arraystretch}{1.5}
\begin{center}
\begin{tabular}
{c@{\hspace{0.6cm}}
S[table-format=-1.3e-1]@{\hspace{0.6cm}}
S[table-format=-1.3e-1]@{\hspace{0.6cm}}
S[table-format=-1.3e-1]}
\toprule
$k$ &
{$a^\CM_{k,0}$} &
{$a^\CM_{k,1}$} &
{$a^\CM_{k,2}$} \\
\hline
1 & 4.870e-3 & -8.936e-4 & 7.854e-4 \\
2 & 7.722e-1 & 5.984e-3  & 5.586e-4 \\
\toprule
$k$ &
{$\omega^\CM_{k,0}\,[E_h/\hbar]$} &
{$\omega^\CM_{k,1}\,[E_h/\hbar]$} &
{$\omega^\CM_{k,2}\,[E_h/\hbar]$}
\\
\hline
1 & 1.289e-1  & -4.571e-3  & 9.421e-4  \\
2 & 3.129e-1  & 6.405e-4  & -6.527e-4  \\
\toprule
$k$ & 
{$\gamma^\CM_{k,0}\,[E_h/\hbar]$} & 
{$\gamma^\CM_{k,1}\,[E_h/\hbar]$} & 
{$\gamma^\CM_{k,2}\,[E_h/\hbar]$}
\\
\hline
1 & 1.875e-2  & 9.274e-4  & 1.651e-3  \\
2 & 9.742e-2  & 4.814e-2  & -1.518e-2  \\
\toprule
$k$ & 
{$\gamma^{\prime\CM}_{k,0}\,[E_h/\hbar]$} & 
{$\gamma^{\prime\CM}_{k,1}\,[E_h/\hbar]$} & 
{$\gamma^{\prime\CM}_{k,2}\,[E_h/\hbar]$}
\\
\hline
1 & 1.387e-1  & 1.161e-2  & -1.543e-2  \\
2 & 9.505e-2  & 5.607e-2  & -1.948e-2  \\
\toprule
\end{tabular}
\end{center}
\end{table}

Standard Bethe-Salpeter or TDDFT calculations of the optical absorption spectra
of solids do not include any radiative reaction forces, and the resulting
oscillator strengths are purely real \cite{Ul2012}. To formally justify
the parameter $\gamma_k'$ in Eq.~\eqref{master} 
one needs an {\em ab initio} approach in
which the dynamics of the electrons and the photon field are coupled and
treated on an equal footing, either at the classical level using Maxwell's
equations, or using QED.  For the latter case, a coupled QED-TDDFT framework
has been developed in the past few 
years~\cite{RuMaBa2011,To2013,FlRuApRu2017,FlEtAl2019}.  More relevant for
the context of our work, Sch\"afer and Johansson~\cite{ScJo2022} recently
proposed a TDDFT formalism that includes dissipation due to classical
Abraham-Lorentz-type radiative reaction forces, and presented applications to
plasmonic systems. 
The functional forms employed here are consistent 
with the mechanisms underlying the DFT, TDDFT, Bethe-Salpeter
and QED-TDDFT frameworks employed in the 
investigations which lead toward an {\em ab initio} understanding 
of the dielectric function.
While at present, to our knowledge, there exist no first-principles
calculations of the optical spectra of periodic solids including classical
radiative reaction forces or QED; given the progress in the field, such results
may emerge in the near future.

In Refs.~[\onlinecite{OuCa2003}] and~[\onlinecite{LaDKJe2010pra}],
inspired by the Clausius–Mossotti relation, the dielectric ratio
\begin{equation}
\label{defrho}
\rho(T_\Delta,\omega) = 
\frac{\epsilon(T_\Delta,\omega)-1}{\epsilon(T_\Delta,\omega)+2}
\doteq f(T_\Delta, \omega)
\end{equation}
was fitted to the Lorentz–Dirac functional form, for $\alpha$-quartz. 
[The fitting to the functional form is indicated by the 
$\doteq$ sign.] In this
paper, we propose to combine the advantages of the approaches 
outlined in 
Refs.~[\onlinecite{OuCa2003,DeJo2012,PrKa2014,ChBaJu2019}],
namely, the inclusion of radiation reaction, and the advantage of
the approach chosen in 
Refs.~[\onlinecite{OuCa2003}] and~[\onlinecite{LaDKJe2010pra}],
which is the dense-material effect encoded
in the Clausius–Mossotti relation. To this end, we first take experimental data
for the temperature-dependent dielectric function of silicon, on the basis of
which we calculate the dielectric ratio $\rho(T_\Delta,\omega)$, which we then
fit with $f(T_\Delta,\omega)$.  The resulting expression for the dielectric
function is
\begin{equation}
\label{CMfit}
\epsilon_{\rm CM}(T_\Delta,\omega)  \doteq
\frac{1 + 2f(T_\Delta,\omega)}{1-f(T_\Delta,\omega)} \,.
\end{equation}
This will be referred to as the Clausius-Mossotti (CM) fit.
The more direct fit
\begin{equation}
\label{LDfit}
\epsilon_{\rm LD}(T_\Delta,\omega) - 1 \doteq f(T_\Delta,\omega)
\end{equation}
will be referred to as the Lorentz-Dirac (LD) fit. 
The quantity
\begin{equation}
\label{deltaeps}
\delta \epsilon(T_\Delta, \omega) =
\left| \epsilon_{\CM}(T_\Delta, \omega) -
\epsilon_{\LD}(T_\Delta, \omega) \right|
\end{equation}
measures the dependence of the fitted dielectric function
on the fitting procedure. Of course,
the quantity $\delta \epsilon(T_\Delta, \omega)$ 
does not include experimental uncertainty
pertaining to the input 
data~\cite{Je1992,Gr2008,ScEtAl2015,VuEtAl1993,SiHoHu1998}.
For the statistical uncertainties 
of the fitting parameters of our model, 
we refer to Tables~I and~II
of the Ref.~[\onlinecite{MoEtAl2022suppl}].

A brief discussion is in order.
While one could argue that the CM fit is favored by physical 
considerations~\cite{LaDKJe2010pra},
one should note the lack
of experimental uncertainty estimates in the 
input data provided in Refs.~[\onlinecite{VuEtAl1993,SiHoHu1998,Gr2008}].
As a byproduct of the alternative fits 
$\rho(T_\Delta, \omega) \approx f(T_\Delta, \omega)$
and $\epsilon(T_\Delta, \omega) - 1 \approx f(T_\Delta, \omega)$,
we are able to estimate the uncertainty 
on the basis of Eq.~\eqref{deltaeps}.
Two remarks are in order.
{\em (i)} Both $\epsilon_{\CM}(T_\Delta, \omega)$ as well as
$\epsilon_{\LD}(T_\Delta, \omega)$
fulfill the Kramers-Kronig relationships
For $\epsilon_{\LD}(T_\Delta, \omega)$, this has been 
shown explicitly in Ref.~[\onlinecite{PrKa2014}],
while for $\epsilon_{\CM}(T_\Delta, \omega)$,
this follows from the relations
$\Re [\epsilon_{\CM}(T_\Delta, \omega)] =
\Re [\epsilon_{\CM}(T_\Delta, -\omega)]$ and
$\Im [\epsilon_{\CM}(T_\Delta, \omega)] = 
- \Im [\epsilon_{\CM}(T_\Delta, -\omega)]$.
These relations allow us to invoke the 
formalism outlined in Sec.~6.6 of Ref.~[\onlinecite{Je2017book}].
{\em (ii)} In the asymptotic limit of large $\omega$, in view of
the limiting process 
\begin{equation}
\label{epsasymp}
\rho(T_\Delta, \omega) = 
\frac{\epsilon(T_\Delta, \omega)-1}{\epsilon(T_\Delta, \omega)+2} 
\;\;
\mathop{\longrightarrow}^{\omega \to \infty} 
\;\;
\frac{\epsilon(T_\Delta, \omega)-1}{3} \,,
\end{equation}
for $\omega\to \infty$, one has
$\epsilon(T_\Delta, \omega) 
\longrightarrow  1 + 3 \rho(T_\Delta, \omega)$.
So, in the asymptotic limits, 
the two fitted functional forms $f(T_\Delta, \omega)$
become equivalent up to the addition of unity, 
and a multiplicative factor three.

\begin{figure*}[t]
\begin{center}
\begin{minipage}{0.9\linewidth}
\begin{center}
\includegraphics[width=0.9\linewidth]{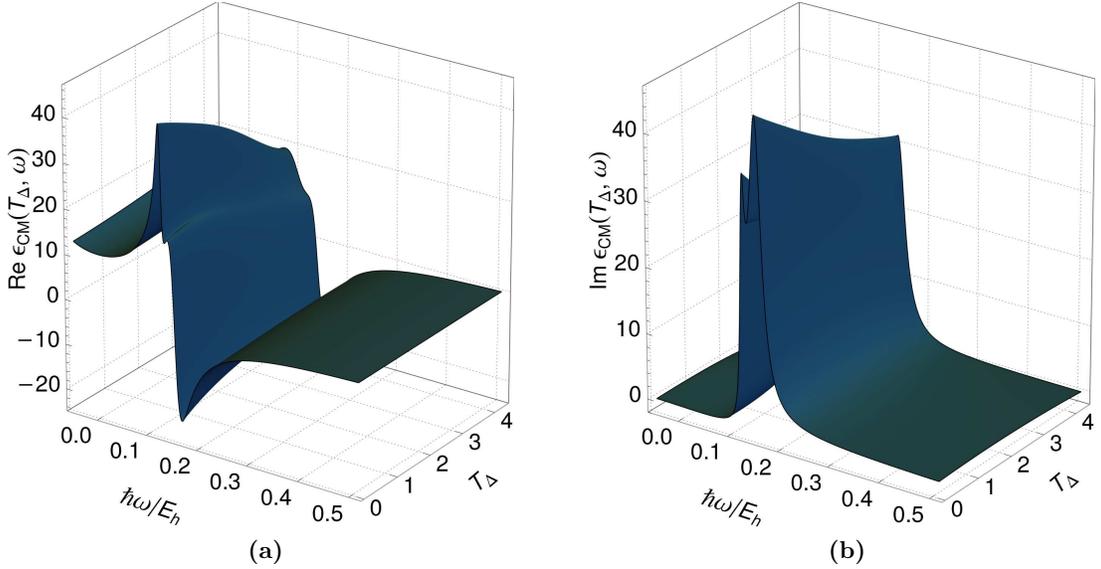}
\caption{\label{fig2} 
Real (a) and imaginary (b) parts of the Clausius-Mossotti
dielectric function $\epsilon_\CM(T_\Delta, \omega)$ as described by 
Eq.~\eqref{CMfit} are plotted as functions of
the reduced temperature $T_\Delta$ and 
driving frequency $\omega$,
for monocrystalline silicon 
with parameters given in
Eq.~\eqref{eq:coefTempCM} and coefficients in Table~\ref{table3}.
This plot shows agreement between the CM fit shown here and 
the LD fit shown in Fig.~\ref{fig4}.}
\end{center}
\end{minipage}
\end{center}
\end{figure*}

%
%
\subsection{Clausius-Mossotti Model}
\label{sec2B}

In the context of the current investigation,
our aim is to find a simple and 
consistent fit to the dielectric function of monocrystalline
(intrinsic) silicon. 
In the current section, we attempt to fit the dielectric 
ratio $\rho(T_\Delta, \omega)$ defined in Eq.~\eqref{defrho}
to the master function given in Eq.~\eqref{master}.
Let us first recall a few essential formulas.
We denote the real and imaginary parts of the complex index of
refraction by $n(\omega)$ and $k(\omega)$, respectively.
These quantities are related to each other by 
the Kramers-Kronig relations (see, e.g., 
Chap.~6 of Ref.~[\onlinecite{Je2017book}]).
The same is true for the real
and imaginary parts of the dielectric function 
$\epsilon(\omega)$, which is given as
$\epsilon(\omega) = [n(\omega) + \ii \, k(\omega)]^2$.
We introduce a phenomenological 
description of the dielectric function by 
simply assuming a 
temperature dependence of the 
individual parameters in Eq.~\eqref{eq:coefTemp}.
This approach has been taken in 
Refs.~[\onlinecite{NISTSPECTRA2014,SiHoHu1998,Ar2007phd}]. 
We thus write the temperature-dependent dielectric
functions as in Eq.~\eqref{defrho},
\begin{multline}
\label{cCMFit}
\rho(T_\Delta,\omega) =
\frac{\epsilon(T_\Delta,\omega) - 1}{\epsilon(T_\Delta,\omega) + 2} 
\\
\approx \rho_\CM(T_\Delta,\omega)
= \frac{\epsilon_\CM(T_\Delta,\omega) - 1}{\epsilon_\CM(T_\Delta,\omega) + 2} 
\\
= \sum_{k=1}^{k_{\rm max}}
\frac{a^\CM_k(T_\Delta) \, [ \, \omega^\CM_k(T_\Delta) \, ]^2 - 
\ii \, \gamma^{\prime\CM}_k(T_\Delta) \, \omega)}
{[ \, \omega^\CM_k(T_\Delta) \, ]^2 
- \omega^2
- \ii \, \omega \, \gamma^\CM_k(T_\Delta)} \,,
\end{multline}
where $k$ counts the number of resonances
and we have indicated the place where we employ the
fitting procedure
by the ``$\approx$'' sign.

\begin{figure*}[t!]
\begin{center}
\includegraphics[width=1.99\columnwidth]{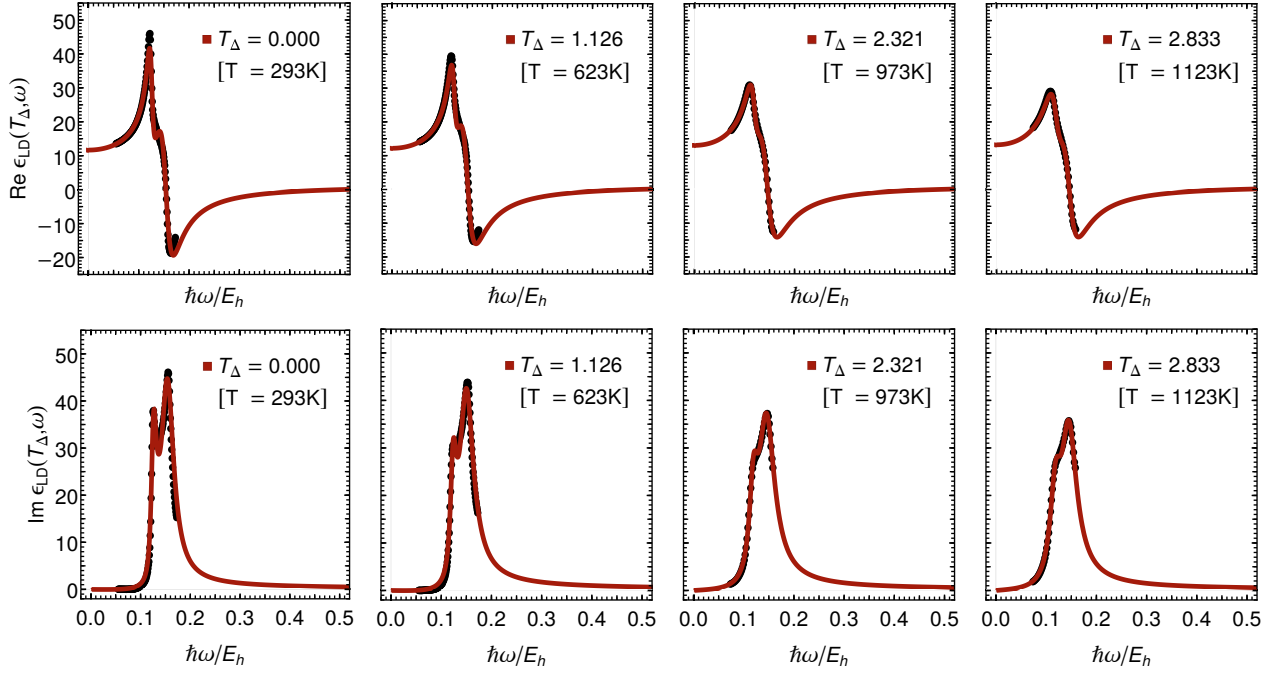}
\end{center}
\caption{\label{fig3}
We present the analog of Fig.~\ref{fig1} for the 
LD as opposed to the CM fitting procedure.
Again, real (top row) and imaginary (bottom row) parts of the dielectric
function $\epsilon_\LD(T_\Delta, \omega)$ 
are plotted as functions of frequency $\omega$, for
monocrystalline silicon for various temperatures,
but here, for the LD fitting procedure.  Experimental data 
\cite{ScEtAl2015,Je1992} (dotted) are found to be in agreement with 
the LD fit (red) defined in Eq.~\eqref{LDfit} 
for temperature-dependent parameters given in
Eq.~\eqref{eq:coefTemp} and coefficients in Table~\ref{table6}. 
For the LD fit, eight additional temperatures, 
between $T_\Delta = 0$ and $T_\Delta = 2.83$, namely, the values
$T_\Delta = 0.273$, $0.444$, $0.614$, $0.785$, $1.126$, $1.297$, 
$1.468$, and $1.638$,
are considered in Ref.~[\onlinecite{MoEtAl2022suppl}].}
\end{figure*}

The temperature-dependent 
real and imaginary parts of $\rho_\CM(T_\Delta,\omega)$ 
can thus be written as follows,
\begin{multline}
\label{eq:reFitCM}
\mathrm{Re}[\rho_\CM(T_\Delta, \omega)] = 
\sum_{k=1}^{k_{\rm max}} a^\CM_k(T_\Delta) 
\\
\times
\frac{ \omega^2 \left[ \, \gamma^\CM_k(T_\Delta) \, 
\gamma_k^{\prime\CM}(T_\Delta) -
[\omega^\CM_k(T_\Delta)]^2 \, \right] +\omega^4 }%
{(\omega^2-[\omega^\CM_k(T_\Delta)]^2)^2+ \omega^2\, 
\left[\gamma^\CM_k(T_\Delta)\right]^2}\,,
\end{multline}
while the imaginary part is
\begin{multline}
\label{eq:imFitCM}
\mathrm{Im}[\rho_\CM(T_\Delta, \omega)] = 
\sum_{k=1}^{k_{\rm max}} a^\CM_k(T_\Delta) \, \omega \, 
\\
\times \frac{ \omega^2 \gamma_k^{\prime\CM}(T_\Delta)
+ \left\{ \gamma^\CM_k(T_\Delta) -
\gamma_k^{\prime\CM}(T_\Delta) \right\} [\omega^\CM_k(T_\Delta)]^2 }%
{ \{ \omega^2-[\omega^\CM_k(T_\Delta)]^2 \}^2 + 
\omega^2\, \left[ \, \gamma^\CM_k(T_\Delta) \, \right]^2}.
\end{multline}
Following Refs.~[\onlinecite{BaPi1979,IcPlWo1976}],
the coefficients $a_k(T_\Delta)$, $\omega_k(T_\Delta)$, 
and $\gamma_k(T_\Delta)$ 
are approximated by quadratic functions in the temperature,
\begin{subequations}
\label{eq:coefTempCM}
\begin{align}
a_k(T_\Delta) =& \;
a^\CM_{k,0} 
+ a^\CM_{k,1} \; T_\Delta 
+ a^\CM_{k,2} \; (T_\Delta)^2 \,,
\\
\omega_k(T_\Delta) = & \;
\omega^\CM_{k,0} 
+ \omega^\CM_{k,1} \; T_\Delta 
+ \omega^\CM_{k,2} \; (T_\Delta)^2 \,,
\\
\gamma_k(T_\Delta) = & \;
\gamma^\CM_{k,0} 
+ \gamma^\CM_{k,1} \; T_\Delta 
+ \gamma^\CM_{k,2} \; (T_\Delta)^2 \,,
\\
\gamma_k^\prime(T_\Delta) = & \;
\gamma^{\prime\CM}_{k,0} 
+ \gamma^{\prime\CM}_{k,1} \; T_\Delta 
+ \gamma^{\prime\CM}_{k,2} \; (T_\Delta)^2 \,,
\end{align}
\end{subequations}
where $T_0=293$K. 
In Tables~\ref{table1} and~\ref{table2}, we show the 
Clausius-Mossotti dielectric ratio coefficients for 
intrinsic silicon, obtained by fitting data 
taken from Refs.~[\onlinecite{VuEtAl1993, SiHoHu1998}]
to Eqs.~\eqref{eq:reFitCM} and~\eqref{eq:imFitCM},
for the first two resonances. 
We find that fits with $k_{\rm max} = 2$ 
lead to satisfactory results.
Coefficients from Tables~\ref{table1} and~\ref{table2} 
are then fitted by assuming a quadratic temperature dependence 
according to Eq.~\eqref{eq:coefTempCM}, 
to obtain a dielectric function for silicon 
$\epsilon(T_\Delta,\omega)$ which is a function of temperature and 
frequency. The coefficients from this fit are given in Table~\ref{table3}. 
In Fig.~\ref{fig1},
the CM fit defined in Eq.~\eqref{CMfit} is plotted alongside 
experimental data\cite{VuEtAl1993, SiHoHu1998}, using the 
coefficients from Table~\ref{table3} and temperature-dependent parameters 
given in Eq.~\eqref{eq:coefTempCM}. 
The similarity of the plots from Figs.~\ref{fig1} and~\ref{fig3}
indicates that the CM fit and the LD fit both
accurately reproduce the experimental data using different methods,
demonstrating that the conclusions of Ref.~[\onlinecite{OuCa2003}]
are more generally applicable.
A unified three-dimensional representation 
for the dielectric function of silicon is given in Fig.~\ref{fig2}.
Based on fits of the functional form given in Eq.~\eqref{eq:coefTempCM},
we obtain the fits presented in Fig.~\ref{fig1},
for individual temperatures. 
A unified three-dimensional representation 
for the real and imaginary parts of the 
temperature- and frequency-dependent dielectric
function $\epsilon_{\CM}$ for silicon
is~given~in~Fig.~\ref{fig2}.

\begin{table}[t]
\caption{\label{table4}
Coefficients resulting from the LD fit,
according to Eqs.~\eqref{eq:reFit} and~\eqref{eq:imFit},
are given for the first resonance of
monocrystalline silicon 
over a range of temperatures 
$0 < T_\Delta < 2.83$.
We recall that $E_h$ is the Hartree energy and 
$\hbar$ is Planck's constant.}
\renewcommand{\arraystretch}{1.25}
\begin{center}
\begin{minipage}{1.00\linewidth}
\begin{tabular}
{c@{\hspace{0.06\linewidth}}
c@{\hspace{0.06\linewidth}}
c@{\hspace{0.06\linewidth}}
c@{\hspace{0.06\linewidth}}c}
\toprule\\[-1em]
$T_\Delta$ &
$a_1^\LD$ &
$\omega_1^\LD\,[E_h/\hbar]$ &
$\gamma_1^\LD\, [E_h/\hbar]$ &
$\gamma^{\prime\LD}_1\, [E_h/\hbar]$ \\[1ex]
\hline
0.000 & 2.817 & 0.1254 & 0.01243 & 0.05791 \\
0.273 & 2.713 & 0.1248 & 0.01305 & 0.05348 \\
0.444 & 2.674 & 0.1245 & 0.01335 & 0.05094 \\
0.614 & 2.568 & 0.1241 & 0.01348 & 0.04748 \\
0.785 & 2.899 & 0.1240 & 0.01545 & 0.04800 \\
0.956 & 2.790 & 0.1236 & 0.01508 & 0.04536 \\
1.126 & 2.643 & 0.1233 & 0.01448 & 0.04078 \\
1.297 & 2.607 & 0.1227 & 0.01491 & 0.03939 \\
1.468 & 2.813 & 0.1224 & 0.01610 & 0.03600 \\
1.638 & 2.967 & 0.1212 & 0.01794 & 0.03360 \\
2.321 & 3.164 & 0.1194 & 0.02062 & 0.02098 \\
2.833 & 4.423 & 0.1190 & 0.02750 & 0.01212 \\
\hline
\hline
\end{tabular}
\end{minipage}
\end{center}
\end{table}

\begin{table}[t]
\caption{\label{table5}
We present the analog of Table~\ref{table4}
for the second resonance of
monocrystalline silicon.
Coefficients resulting 
from the Lorentz-Dirac fit,
according to Eqs.~\eqref{eq:reFit} and~\eqref{eq:imFit},
are given over a range of temperatures 
$0 < T_\Delta < 2.83$.}
\renewcommand{\arraystretch}{1.25}
\begin{center}
\begin{minipage}{1.0\linewidth}
\begin{tabular}
{c@{\hspace{0.06\linewidth}}
c@{\hspace{0.06\linewidth}}
c@{\hspace{0.06\linewidth}}
c@{\hspace{0.06\linewidth}}c}
\toprule\\[-1em]
$T_\Delta$ &
$a_2^\LD$ &
$\omega_2^\LD\,[E_h/\hbar]$ &
$\gamma_2^\LD\, [E_h/\hbar]$ &
$\gamma^{\prime\LD}_2\, [E_h/\hbar]$ \\[1ex]
\hline
0.000 & 7.844 & 0.1545 & 0.02994 & 0.00792 \\
0.273 & 8.245 & 0.1529 & 0.03115 & 0.01272 \\
0.444 & 8.325 & 0.1520 & 0.03151 & 0.01589 \\
0.614 & 8.588 & 0.1520 & 0.03247 & 0.01768 \\
0.785 & 8.344 & 0.1515 & 0.03245 & 0.01880 \\
0.956 & 8.410 & 0.1507 & 0.03275 & 0.02136 \\
1.126 & 8.537 & 0.1501 & 0.03284 & 0.02421 \\
1.297 & 8.749 & 0.1495 & 0.03407 & 0.02427 \\
1.468 & 8.601 & 0.1490 & 0.03422 & 0.02528 \\
1.638 & 8.709 & 0.1475 & 0.03592 & 0.02379 \\
2.321 & 8.873 & 0.1455 & 0.03825 & 0.02284 \\
2.833 & 7.789 & 0.1453 & 0.03648 & 0.02526 \\
\hline
\hline
\end{tabular}
\end{minipage}
\end{center}
\end{table}

\begin{table}[t]
\caption{\label{table6}
Coefficients associated with 
the Lorentz-Dirac model given in
Eq.~(\ref{eq:coefTemp}), found from fitting 
the coefficients in Table~\ref{table4} and~Table~\ref{table5}
as functions of temperature, are given
for monocrystalline silicon.}
\renewcommand{\arraystretch}{1.5}
\begin{center}
\begin{tabular}
{c@{\hspace{0.6cm}}
S[table-format=-1.3e-1]@{\hspace{0.6cm}}
S[table-format=-1.3e-1]@{\hspace{0.6cm}}
S[table-format=-1.3e-1]}
\toprule
$k$ & 
{$a^\LD_{k,0}$} & 
{$a^\LD_{k,1}$} & 
{$a^\LD_{k,2}$}
\\
\hline
1& {$2.892\times 10^0$} & -6.339e-1  & 3.890e-1  \\
2& {$7.864\times 10^0$} & {$1.121\times 10^0$}  & -3.754e-1  \\
\toprule
$k$ & 
{$\omega^\LD_{k,0}\,[E_h/\hbar]$} & 
{$\omega^\LD_{k,1}\,[E_h/\hbar]$} & 
{$\omega^\LD_{k,2}\,[E_h/\hbar]$}
\\
\hline
1& 1.255e-1  & -2.154e-3  & -9.91e-5  \\
2& 1.544e-1  & -4.408e-3  & 3.685e-4  \\
\toprule
$k$ & 
{$\gamma^\LD_{k,0}\,[E_h/\hbar]$} & 
{$\gamma^\LD_{k,1}\,[E_h/\hbar]$} & 
{$\gamma^\LD_{k,2}\,[E_h/\hbar]$}
\\
\hline
1& 1.308e-2  & -3.701e-4  & 1.819e-3  \\
2& 2.980e-2  & 4.011e-3  & -4.602e-4  \\
\toprule
$k$ & 
{$\gamma'^\LD_{k,0}\,[E_h/\hbar]$} & 
{$\gamma'^\LD_{k,1}\,[E_h/\hbar]$} & 
{$\gamma'^\LD_{k,2}\,[E_h/\hbar]$}
\\
\hline
1& 5.732e-2  & -1.261e-2  & -1.211e-3  \\
2& 8.672e-3  & 1.688e-2  & -4.117e-3  \\
\toprule
\end{tabular}
\end{center}
\end{table}

%
%
\subsection{Lorentz-Dirac Model}
\label{sec2C}

\begin{figure*}[t]
\begin{center}
\begin{minipage}{0.9\linewidth}
\begin{center}
\includegraphics[width=0.9\linewidth]{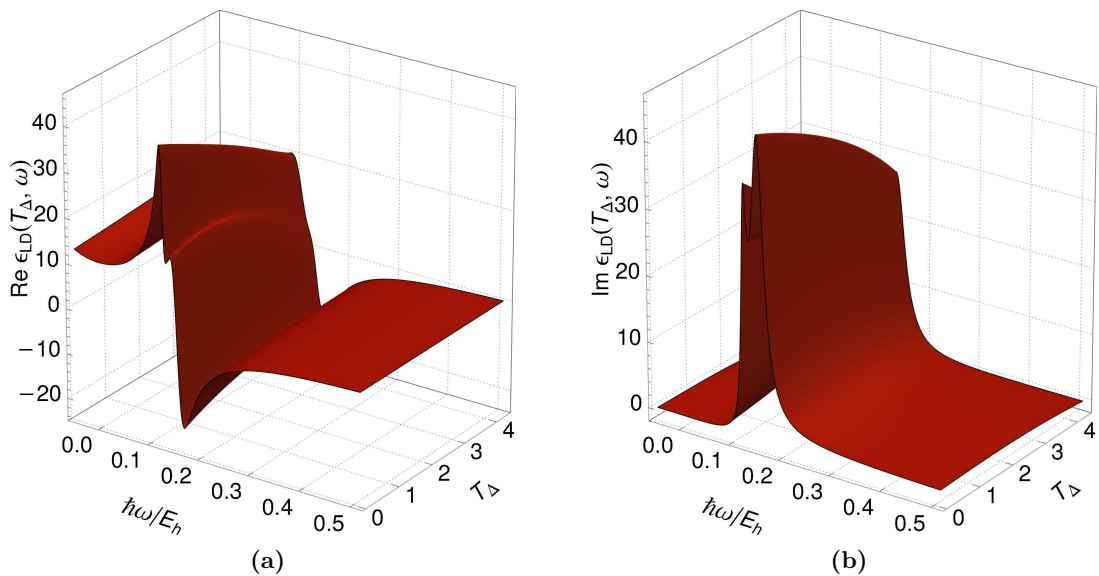}
\end{center}
\caption{\label{fig4} 
We present the analog of Fig.~\ref{fig2} for the LD fit,
as opposed to the CM fit of the dielectric function.
Real (a) and imaginary (b) parts of the dielectric function 
$\epsilon_\LD(T_\Delta, \omega)$, as described by Eq.~\eqref{LDfit},
are plotted as functions of
the reduced temperature $T_\Delta$ and driving frequency $\omega$,
for monocrystalline silicon with parameters given in
Eq.~\eqref{eq:coefTemp} and coefficients in Table~\ref{table6}.
We find good agreement of the LD fit shown here and 
the CM fit shown in Fig.~\ref{fig2}.}
\end{minipage}
\end{center}
\end{figure*}

As discussed in Sec.~\ref{sec2A}, 
we now turn to the second method of fitting the 
dielectric function of the reference substrate, 
monocrystalline silicon, which is based on 
a direct fit of the experimentally
determined dielectric function $\epsilon(T_\Delta, \omega)$ to the 
Lorentz-Dirac master function Eq.~\eqref{master}
with free parameters.
We introduce a phenomenological 
description of the dielectric function by 
assuming a temperature dependence of the 
individual parameters,
and write the temperature-dependent dielectric
function in terms of a functional form 
inspired by the master function given in Eq.~\eqref{master},
but with temperature-dependent parameters,
\begin{multline}
\label{cLDFit}
\epsilon(T_\Delta,\omega) \approx 
\epsilon_{\LD}(T_\Delta,\omega) = 1 
\\
+ \sum_{k=1}^{k_{\rm max}}
\frac{a^\LD_k(T_\Delta) \, \left\{ [ \, \omega^\LD_k(T_\Delta) \, ]^2 -
\ii \, \gamma^{\prime\LD}_k(T_\Delta) \, \omega \right\} }%
{[ \, \omega^\LD_k(T_\Delta) \, ]^2
- \omega^2
- \ii \, \omega \, \gamma^\LD_k(T_\Delta) } \,,
\end{multline}
where we employ a fitting procedure during the step 
that is marked with the $\approx$ sign.
The temperature-dependent 
real part of $\epsilon(T_\Delta,\omega)$ 
can thus be written as follows,
\begin{multline}
\label{eq:reFit}
\mathrm{Re}[\epsilon_{\LD}(T_\Delta, \omega)] =
1 + \sum_{k=1}^{k_{\rm max}} a^\LD_k(T_\Delta)
\\
\times
\frac{ \omega^2 \left[ \, \gamma^\LD_k(T_\Delta) \,
\gamma_k^{\prime\LD}(T_\Delta) -
\omega^\LD_k(T_\Delta)^2 \, \right] +\omega^4 }%
{(\omega^2-[\omega^\LD_k(T_\Delta)]^2)^2+ \omega^2\,
\left[\gamma^\LD_k(T_\Delta)\right]^2} \,.
\end{multline}
The imaginary part is given as follows,
\begin{multline}
\label{eq:imFit}
\mathrm{Im}[\epsilon_{\LD}(T_\Delta,\omega)]
= \sum_{k=1}^{k_{\rm max}} a^\LD_k(T_\Delta) \, \omega 
\\
\times \frac{ \omega^2 \gamma_k^{\prime\LD}(T_\Delta)
+ \left\{ \gamma^\LD_k(T_\Delta) -
\gamma_k^{\prime\LD}(T_\Delta) \right\} [\omega^\LD_k(T_\Delta)]^2 }%
{ \{ \omega^2-[\omega_k(T_\Delta)]^2 \}^2 +
\omega^2\, \left[ \, \gamma^\LD_k(T_\Delta) \, \right]^2}.
\end{multline}
In full analogy with the
approach outlined in
Refs.~[\onlinecite{BaPi1979,IcPlWo1976}]
and in Sec.~\ref{sec2B} [see Eq.~\eqref{eq:coefTempCM}],
the coefficients $a_k(T_\Delta)$, $\omega_k(T_\Delta)$,
and $\gamma_k(T_\Delta)$
are approximated by quadratic functions in the temperature,
\begin{subequations}
\label{eq:coefTemp}
\begin{align}
a_k(T_\Delta) = & \;
a^\LD_{k,0} 
+ a^\LD_{k,1} \; T_\Delta 
+ a^\LD_{k,2} \; (T_\Delta)^2\,,
\\
\omega_k(T_\Delta) = & \;
\omega^\LD_{k,0} 
+ \omega^\LD_{k,1} \; T_\Delta 
+ \omega^\LD_{k,2} \; (T_\Delta)^2\,,
\\
\gamma_k(T_\Delta) = & \;
\gamma^\LD_{k,0} 
+ \gamma^\LD_{k,1} \; T_\Delta 
+ \gamma^\LD_{k,2} \; (T_\Delta)^2\,,
\\
\gamma_k^\prime(T_\Delta) = & \;
\gamma^{\prime\LD}_{k,0} 
+ \gamma^{\prime\LD}_{k,1} \; T_\Delta 
+ \gamma^{\prime\LD}_{k,2} \; (T_\Delta)^2 \,,
\end{align}
\end{subequations}
where $T_0=293$K. 
In Tables~\ref{table4} and~\ref{table5}, we show the 
Clausius-Mossotti dielectric ratio coefficients for 
intrinsic silicon, obtained by fitting data 
taken from Refs.~[\onlinecite{VuEtAl1993, SiHoHu1998}]
to Eqs.~\eqref{eq:reFit} and~\eqref{eq:imFit},
for the first two resonances. 
Coefficients from Tables~\ref{table4} and~\ref{table5} 
are then fitted by assuming a quadratic temperature dependence 
according to Eq.~\eqref{eq:coefTemp}, to obtain the
dielectric function for silicon $\epsilon(T_\Delta,\omega)$ 
as a function of temperature and 
driving frequency. The coefficients from this fit 
are given in Table~\ref{table6}. The accuracy of the fits can be seen in Fig.~\ref{fig3} 
where the LD fit defined in Eq.~\eqref{LDfit} is plotted alongside 
experimental data\cite{VuEtAl1993, SiHoHu1998}, using the 
coefficients from Table~\ref{table6} and temperature-dependent parameters 
given in Eq.~\eqref{eq:coefTemp}. 
The similarity of the plots given in Fig.~\ref{fig3} 
to those in Fig.~\ref{fig1}
reveals that the CM fitting and the LD fitting both
accurately reproduce the experimental data using different methods.
A unified three-dimensional representation 
for the dielectric function of silicon is given in Fig.~\ref{fig4}.

For the LD fits, one particular point
is worth mentioning: If one
searches for the best fit parameters (in the sense
of a least-squares approach) for the
fitting procedure $\epsilon_\LD(T_\Delta, \omega)
\doteq 1 + f(T_\Delta, \omega)$, with unrestricted
fit parameters $a_k$, $\omega_k$, $\gamma_k$,
and $\gamma'_k$, then one may incur, for certain temperatures,
fitting functions for $\epsilon_\LD(T_\Delta, \omega)$
whose imaginary part, for small and positive $\omega$,
turns slightly negative.
This behavior is unphysical. We have therefore
implemented the condition 
\begin{multline}
\left.
\frac{\partial}{\partial \omega}
\mathrm{Im}[\epsilon_{\LD}(T_\Delta,\omega)]
\right|_{\omega = 0} = \\
\sum_{k=1}^{k_{\rm max}} 
\frac{ a^\LD_k(T_\Delta) }{ [\omega^\LD_k(T_\Delta)]^2 }
\left( \gamma_k^{\LD}(T_\Delta)
- \gamma'^\LD_k(T_\Delta) \right) \;
 >\; 0
\end{multline}
in the nonlinear fitting procedure~\cite{Wo1999}
via additional derivatives.
We observe that, as we ensure that the first
derivative of the fitted imaginary part at zero frequency
is forced to be positive, the entire fitted imaginary 
part consistently assumes positive values over the 
entire frequency range $0 < \omega < \infty$.

\begin{table}[t]
\caption{\label{table7} 
Short-distance coupling parameters $C_3^\CM$ and $C_3^\LD$ of the
Casimir-Polder potential are given for helium 
interacting with a monocrystalline silicon surface.
Results are given for the CM and the LD fitting procedures given 
in Eqs.~\eqref{CMfit} and Eq.~\eqref{LDfit}, respectively.
The numerical values are obtained using the 
integrals given in Eqs.~\eqref{C3CM} and~\eqref{C3LD}.
The data is plotted in Fig.~\ref{fig5}, as a function of the 
temperature.}
\renewcommand{\arraystretch}{1}
\begin{center}
\begin{minipage}{0.95\linewidth}
\begin{tabular}
{c@{\hspace{0.6cm}}c@{\hspace{0.6cm}}c@{\hspace{0.6cm}}c@{\hspace{0.15cm}}c}
\hline
\hline
\multicolumn{4}{c}{\stdrule%
Helium on Silicon} & \\
\multicolumn{4}{c}{\stdrule%
Short-Range $C_3$ Coefficient} & \\
\stdrule%
$T_\Delta$ &
$C_3^\CM$ [$a_0^3 E_h$] &
$C_3^\LD$ [$a_0^3 E_h$] &
$\%$ difference\\ 
\hline
0.000 &	0.04906 & 0.04950 & 0.91 \\
0.273 &	0.05013 & 0.05068 & 1.10 \\
0.444 &	0.05128 & 0.05132 & 0.08 \\
0.614 &	0.05143 & 0.05173 & 0.57 \\
0.785 &	0.05198 & 0.05253 & 1.04 \\
0.956 &	0.05247 & 0.05276 & 0.54 \\
1.126 &	0.05295 & 0.05297 & 0.04 \\
1.297 &	0.05321 & 0.05311 & 0.18 \\
1.468 &	0.05376 & 0.05317 & 1.09 \\
1.638 &	0.05254 & 0.05274 & 0.38 \\
2.321 &	0.05066 & 0.05123 & 1.13 \\
2.833 &	0.05022 & 0.05021 & 0.03 \\
\hline
\hline
\end{tabular}
\end{minipage}
\end{center}
\end{table}

\begin{table}[t]
\caption{\label{table8}
Long-distance coupling parameters $C_4^\CM$ and $C_4^\LD$ of the
Casimir-Polder potential are given for helium 
interacting with a monocrystalline silicon surface.
Results are given for the CM and the LD fitting procedures given 
in Eqs.~\eqref{CMfit} and Eq.~\eqref{LDfit}, respectively.
Values are compared with Eqs.~\eqref{C4CM} and~\eqref{C4LD} 
in Fig.~\ref{fig6}.}
\renewcommand{\arraystretch}{1}
\begin{center}
\begin{minipage}{0.95\linewidth}
\begin{tabular}%
{c@{\hspace{0.5cm}}c@{\hspace{0.5cm}}c@{\hspace{0.5cm}}c@{\hspace{0.5cm}}c}
\hline
\hline
\multicolumn{4}{c}{\stdrule%
Helium on Silicon} & \\
\multicolumn{4}{c}{\stdrule%
Long-Range $C_4$ Coefficient} & \\
\stdrule%
$T_\Delta$ &
$C_4^\CM$ [$a_0^4E_h$] &
$C_4^\LD$ [$a_0^4E_h$] &
$\%$ difference\\ 
\hline
0.000 &	15.32 &	15.40 &	0.55 \\
0.273 &	15.37 &	15.49 &	0.76 \\
0.444 &	15.41 &	15.50 &	0.61 \\
0.614 &	15.42 &	15.55 &	0.82 \\
0.785 &	15.44 &	15.57 &	0.84 \\
0.956 &	15.45 &	15.56 &	0.70 \\
1.126 &	15.47 &	15.55 &	0.51 \\
1.297 &	15.49 &	15.60 &	0.70 \\
1.468 &	15.52 &	15.62 &	0.63 \\
1.638 &	15.59 &	15.69 &	0.65 \\
2.321 &	15.67 &	15.78 &	0.69 \\
2.833 &	15.83 &	15.82 &	0.04 \\
\hline
\hline
\end{tabular}
\end{minipage}
\end{center}
\end{table}

%
%
\section{Atom-Surface Potentials}
\label{sec3}

The transition of the atom-surface potential 
from the short-range to the long-range region has 
been discussed at length in the literature
(see, e.g., Refs.~[\onlinecite{LaLi1960vol8,AnPiSt2004,FrJaMe2002,LaDKJe2010pra}]
and references therein). It is well known that 
the atom-surface potentials $V(z)$
mediated by the exchange of virtual photons
change from a $1/z^3$ short-range asymptotic behavior
to a $1/z^4$ long-range asymptotic behavior
($z$ is the atom-wall distance).
The $1/z^3$ short-range asymptotic behavior
persists for $z \ll a_0/\alpha$,
while the $1/z^4$ long-range asymptotic behavior
is relevant for long range, $z \gg a_0/\alpha$,
where $a_0$ is the Bohr radius and $\alpha$ 
is the fine-structure constant.
The asymptotic forms are 
\begin{subequations}
\begin{align}
\label{C3}
V(z) =& \; - \frac{C_3}{z^3} 
= - ( C_3 )_\au \;
\frac{E_h}{(z/a_0)^3} \,,
\quad
a_0 \ll z \ll \frac{a_0}{\alpha} \,,
\\[2ex]
\label{C4}
V(z) =& \; - \frac{C_4}{z^4} 
= - ( C_4 )_\au \;
\frac{E_h}{(z/a_0)^4} \,,
\quad
z \gg \frac{a_0}{\alpha} \,.
\end{align}
\end{subequations}
Here, we denote the numerical value of the $C_3$ and $C_4$ 
coefficients, measured in atomic units, 
by $( C_3 )_\au$ and $( C_4 )_\au$, respectively.

\begin{figure}[t]
\begin{center}
\includegraphics[width=.9\columnwidth]{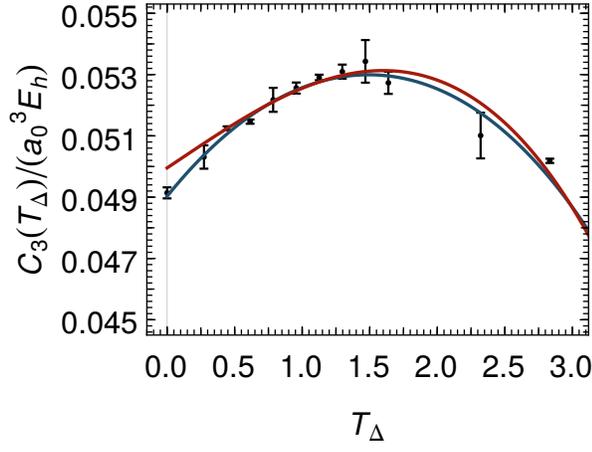}
\end{center}
\caption{\label{fig5}
Short-distance coupling parameters $C_3^\CM(T_\Delta)$ and 
$C_3^\LD(T_\Delta)$ of the Casimir-Polder potential, 
are plotted as functions of $T_\Delta$ for helium interacting 
with a monocrystalline silicon surface. Data points (black dots) are 
taken from Table~\ref{table7}.  The blue curve corresponds to the 
Clausius-Mossotti fit, given in Eq.~\eqref{C3CM}, while 
the red curve corresponds to the Lorentz-Dirac fit, given 
in Eq.~\eqref{C3LD}.  The data points are taken at
$T_\Delta = 0.000$, $0.273$, $0.444$, $0.614$, $0.785$,
$0.956$, $1.126$, $1.297$, $1.468$, $1.638$, $2.321$, 
and $2.833$.}
\end{figure}

\begin{figure}[t]
\begin{center}
\includegraphics[width=.9\columnwidth]{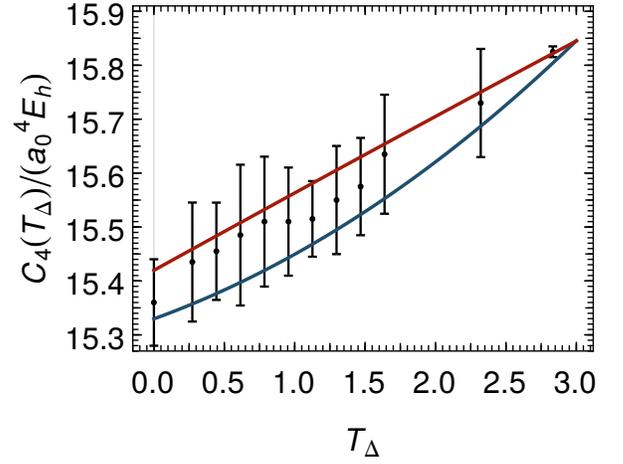}
\end{center}
\caption{\label{fig6}
Long-distance coupling parameters $C_4^\CM(T_\Delta)$ and 
$C_4^\LD(T_\Delta)$ of the Casimir-Polder potential,
are plotted as functions of $T_\Delta$ for helium 
interacting with a monocrystalline silicon surface .
Data points (black dots) are taken from Table~\ref{table8}. 
The blue curve corresponds to the Clausius-Mossotti fit,
given in Eq.~\eqref{C4CM}, while the red curve corresponds to the 
Lorentz-Dirac fit, given in Eq.~\eqref{C4LD}.
The data points are taken at
$T_\Delta = 0.000$, $0.273$, $0.444$, $0.614$, $0.785$,
$0.956$, $1.126$, $1.297$, $1.468$, $1.638$, $2.321$, 
and $2.833$.}
\end{figure}

The $C_3$ and $C_4$ coefficients are, in a natural way,
temperature-dependent, as they depend on the dielectric function
of the substrate 
material~\cite{AnPiSt2004,AnPiSt2005,AnPiStSv2006,AnPiStSv2008}.
Hence, there is a functional 
relationship $C_3 = C_3(T_\Delta)$ and $C_4 = C_4(T_\Delta)$.
It is thus clear that,
for atom-surface interaction studies, it is convenient
to have analytic models for the 
temperature-dependent dielectric function
of a material; we here consider the case
of intrinsic silicon. The 
temperature dependence of the coefficient
$C_3(T_\Delta)$, which governs the short-distance behavior 
of the Casimir-Polder potential, can be written as 
follows~\cite{LaLi1960vol8,AnPiSt2004,FrJaMe2002,LaDKJe2010pra},
\begin{equation}
\label{eq:c3}
C_3(T_\Delta) =
\frac{\hbar}{16\pi^2\epsilon_0}
\int_0^\infty
\dd\omega \; \alpha(\ii\omega) \;
\frac{\epsilon(T_\Delta,\ii\omega)-1}{\epsilon(T_\Delta,\ii\omega)+1} \,.
\end{equation}
Based on the two fitting procedures 
given in Eqs.~\eqref{cCMFit} and~\eqref{cLDFit}, 
we can define the coefficients
\begin{align}
\label{C3CM}
C^\CM_3(T_\Delta) =& \; \frac{\hbar}{16\pi^2\epsilon_0} \int_0^\infty
\dd\omega \; \alpha(\ii\omega) \;
\frac{\epsilon_\CM(T_\Delta,\ii\omega)-1}{\epsilon_\CM(T_\Delta,\ii\omega)+1} \,,
\end{align}
for the Clausius-Mossotti fit, and analogously
\begin{align}
\label{C3LD}
C^\LD_3(T_\Delta) =& \; \frac{\hbar}{16\pi^2\epsilon_0} \int_0^\infty
\dd\omega \; \alpha(\ii\omega) \;
\frac{\epsilon_\LD(T_\Delta,\ii\omega)-1}{\epsilon_\LD(T_\Delta,\ii\omega)+1} 
\end{align}
for the Lorentz-Dirac fit.
In order to calculate $C_3(T_\Delta)$,
we need the dynamic polarizability 
of the atom $\alpha(\omega)$. In 
Appendix~\ref{appB}, we describe 
a rather universally applicable scheme for the 
calculation of the dynamic polarizability 
of an arbitrary atom, based on tabulated
oscillator strength for a limited set of 
transitions, augmented by a matching (at high energy)
against the Thomas-Reiche-Kuhn (TRK) sum rule \cite{ReTh1925zahl,Ku1925}.
The method uses the fact that, at a purely imaginary 
argument, the dynamic polarizability 
is a smooth function which avoids the 
singularities of the integrand 
at the resonance frequencies.
We have applied the method to both 
atomic hydrogen as well as helium, neon, argon, krypton and xenon.
For helium interacting with a silicon 
surface, results are given in Table~\ref{table7}.
Results for helium interacting with silicon are also shown in
Fig.~\ref{fig5}.
A remark is in order.
The numerical results given in Table~\ref{table7} 
correspond to the parameter fits 
for individual temperatures,
as outlined in Eqs.~\eqref{CMfit} and Eq.~\eqref{LDfit},
and in Tables~\ref{table1} and~\ref{table2}.
The smooth curves in Fig.~\ref{fig5} (and analogously in Fig.~\ref{fig6})
constitute the results of 
plotting Eqs.~\eqref{CMfit} and Eq.~\eqref{LDfit}
using the smooth temperature-dependent model 
outlined in the coefficients given in 
Tables~\ref{table3} and~\ref{table4}.
For room temperature, these are in agreement with 
those recently presented in Ref.~[\onlinecite{ZhTaRa2017}].
A separate least-squares fit using a quadratic polynomial in $T_\Delta$
yields the following result for the temperature-dependent
$C_3$ coefficients for helium on silicon,
\begin{align}
C^\CM_3(T_\Delta) \approx & \; 
C^\CM_3(0) \, \left\{ 1 + c_{\rm CM} \, T_\Delta + 
d_{\rm CM}(T_\Delta)^2 \right\} \,,
\end{align}
where $C^\CM_3(0) = 0.04905$,
$c_{\rm CM} = 0.10569$, and $d_{\rm CM} = -0.034338$.
Analogously, one obtains
\begin{align}
C^\LD_3(T_\Delta) \approx & \; 
C^\LD_3(0) \, \left\{ 1 + c_{\rm LD} T_\Delta + d_{\rm LD} (T_\Delta)^2 \right\} \,,
\end{align}
where $C^\LD_3(0) = 0.04923$,
$c_{\rm LD} = 0.09939$, and $d_{\rm LD} = -0.033309$.

For the other atoms under investigation,
the $C_3$ coefficient read as follows.
For H, we obtain, using the Lorentz-Dirac fit, in atomic units, 
at room temperature,
a result of $0.1042$, while for Ne, Ar, Kr, and Xe, the results
for $C_3$, in atomic units, read as
$0.1080$, $0.3914$, $0.5853$ and 
$0.9157$, respectively. 

Let us now investigate the long-range asymptotic 
behavior, as described by Eq.~\eqref{C4},
and let us consider the temperature dependence
of the $C_4$ long-range coefficient.
Indeed, the $C_4$ coefficient governing the 
long-distance behavior $z \gg a_0/\alpha$ of the 
Casimir-Polder potential can be written 
as~\cite{LaLi1960vol8,AnPiSt2004,LaDKJe2010pra}
\begin{equation}
\label{C4_T_Delta}
C_4(T_\Delta)=
\frac{\hbar c}{2\pi}\frac{3}{8} \; \alpha(0) \;
\phi(T_\Delta) \,,
\end{equation}
where $\alpha(0) = \alpha(\omega = 0)$ 
is the static polarizability of the 
atom. The emergence of $\alpha(0)$ in the result illustrates 
the fact that, for large atom-wall separation,
the interaction is mediated by very low-energy virtual
photons. The $\phi$ function in Eq.~\eqref{C4_T_Delta}
is given as an integral, as follows,
\begin{equation}
\label{eq:phi4Fun}
\phi(T_\Delta)=
\int_1^\infty \dd p \, \frac{H(\epsilon(T_\Delta,0),p)}{p^4} \,,
\end{equation}
where the $H$ function reads~\cite{LaLi1960vol8,AnPiSt2004,LaDKJe2010pra}
\begin{equation}
\label{eq:hFun}
H(\epsilon, p)=
\frac{\sqrt{\epsilon-1+p^2}-p}
{\sqrt{\epsilon-1+p^2}+p}
+(1-2p^2) 
\frac{\sqrt{\epsilon-1+p^2}-p \, \epsilon}
{\sqrt{\epsilon-1+p^2}+p \, \epsilon}.
\end{equation}
The $C_4$ coefficient can
be calculated on the basis of the Clausius-Mossotti fit
described in Sec.~\ref{sec2B},
\begin{subequations}
\label{C4CM}
\begin{align}
C^\CM_4(T_\Delta) =& \;
\frac{\hbar c}{2\pi}\frac{3}{8}
\alpha(0) \; \phi_\CM(T_\Delta) \,,
\\[0.1133ex]
\phi_\CM(T_\Delta) =& \;
\int_1^\infty \dd p \, \frac{H(\epsilon_\CM(T_\Delta,0),p)}{p^4} \,,
\end{align}
\end{subequations}
or the Lorentz-Dirac fit,
described in Sec.~\ref{sec2C},
\begin{subequations}
\label{C4LD}
\begin{align}
C_4^\LD(T_\Delta) =& \;
\frac{\hbar c}{2\pi}\frac{3}{8}
\alpha(0) \; \phi_\LD(T_\Delta) \,,
\\[0.1133ex]
\phi_\LD(T_\Delta) =& \;
\int_1^\infty \dd p \, \frac{H(\epsilon_\LD(T_\Delta,0),p)}{p^4} \,.
\end{align}
\end{subequations}
An analytic result for $\phi$, expressed with logarithms,
reads as follows,
\begin{multline}
\label{phires}
\phi(T_\Delta) =
2 \left[
\frac{2 \epsilon(T_\Delta,0)^3-4 \epsilon (T_\Delta,0)^2+3 \epsilon (T_\Delta,0)+1 }%
{4 (\epsilon (T_\Delta,0)-1)^{3/2}} L_1
\right.
\\[0.1133ex]
+ \frac{\epsilon (T_\Delta,0)^2}{2 \sqrt{\epsilon (T_\Delta,0)+1}} \,
\left\{ L_2 - L_3 \right\}
+ \frac{1}{6 (\epsilon(T_\Delta,0)-1)} 
\bigl[ 6 \epsilon (T_\Delta,0)^2 
\\[0.1133ex]
\left. - 3 \epsilon (T_\Delta,0)^{3/2}-4 \epsilon (T_\Delta,0)-
3 \sqrt{\epsilon (T_\Delta,0)}+10
\bigr] \right] \,.
\end{multline}
Here, the logarithmic terms are
\begin{subequations}
\begin{align}
L_1 =& \;
\ln\left(\frac{\sqrt{\epsilon (T_\Delta,0)}-\sqrt{\epsilon (T_\Delta,0)-1}}%
{\sqrt{\epsilon (T_\Delta,0)} + \sqrt{\epsilon (T_\Delta,0)-1}}\right) \,,
\\
L_2 =& \; \ln \left(\frac{\sqrt{\epsilon(T_\Delta,0)+1}-1}%
{\sqrt{\epsilon (T_\Delta,0)+1}+1}\right) \,,
\\
L_3 =& \; 
\ln \left(\frac{\sqrt{\epsilon (T_\Delta,0)+1}-\sqrt{\epsilon (T_\Delta,0)}}
{ \sqrt{\epsilon (T_\Delta,0)+1} + \sqrt{\epsilon (T_\Delta,0)} }\right) \,.
\end{align}
\end{subequations}
Our result for $\phi(T_\Delta)$ is in agreement with the 
result given in Eq.~(23) of Ref.~[\onlinecite{AnPiSt2004}],
but differs in its functional form; we attempt to 
reduce the complexity of the functions involved.
The term in square brackets in Eq.~\eqref{phires}
approximates unity in the limit
of a perfectly conducting surface, $\epsilon(T_\Delta,0) \to \infty$.
The correction terms about the limit 
of large $\epsilon(T_\Delta,0)$ can be expanded in a series 
in inverse half-integer powers of $\epsilon(T_\Delta, 0)$.
The first two correction terms lead to the expression
\begin{multline}
\phi(T_\Delta) = 2 \biggl[ 1 - \frac{5}{4 \sqrt{\epsilon(T_\Delta, 0)}} 
\\[0.1133ex]
+ \frac{22}{15 \epsilon(T_\Delta, 0)}
+ \mathcal{O}\left( 
\frac{\ln[\epsilon(T_\Delta, 0)]}{\epsilon(T_\Delta, 0)^{3/2}}\right)
\biggr] \,.
\end{multline}
Numerical values for $C_4(T_\Delta)$
are calculated for each value of $T_\Delta$,
for both the CM fitting procedure
[according to Eq.~\eqref{C4CM}]
and the LD fitting method
[according to Eq.~\eqref{C4LD}].
Results for helium atoms interacting with 
a silicon surface are given in Table~\ref{table8}
and Fig.~\ref{fig6}.

Conversely, a fit using a quadratic polynomial in $T_\Delta$
yields the following result for the temperature-dependent
$C_4$ coefficients for helium on silicon,
\begin{align}
C^\CM_4(T_\Delta) \approx & \;
C^\CM_4(0) \, \left\{ 1 + f_{\rm CM} \, T_\Delta +
g_{\rm CM}(T_\Delta)^2 \right\} \,,
\end{align}
where $C^\CM_4(0) = 15.32$,
$f_{\rm CM} = 0.0076444$, and $g_{\rm CM} = 0.0014732$.
Analogously, one obtains
\begin{align}
C^\LD_4(T_\Delta) \approx & \;
C^\LD_4(0) \, \left\{ 1 + f_{\rm LD} T_\Delta + g_{\rm LD} (T_\Delta)^2 \right\} \,,
\end{align}
where $C^\LD_4(0) = 15.40$,
$f_{\rm LD} = 0.013560$, and $g_{\rm LD} = -0.0013667$.

The relative discrepancies between the
CM and LD fits for $C_4$
are commensurate with those for the corresponding $C_3$ coefficients,
as listed in Tables~\ref{table7} and~\ref{table8}.
This is consistent with fact that 
the $C_4$ coefficients are 
determined by the static value $\epsilon(T_\Delta, \omega =0)$
of the dielectric function, which can be determined 
to roughly the same accuracy 
as the integral over all frequencies,
which enters Eq.~\eqref{eq:c3}.
Note, also, that the static value of the 
helium polarizability is well known
from Refs.~[\onlinecite{YaBaDaDr1996,PaSa2000}].
A further remark might be in order.
The relative difference of the 
numerical values for the $C_4$ coefficients,
obtained from the CM and LD fits, is smaller 
by about a factor of five than 
the relative difference of the static dielectric 
function $\epsilon(T_\Delta, 0)$,
obtained from either fit.
This somewhat surprising observation finds a natural
explanation when one considers the numerically
small variation of the $\phi$ function
with respect to the value of $\epsilon(T_\Delta, 0)$,
in the relevant range $\epsilon(T_\Delta, 0) \approx 11.5$.
In consequence, the $C_4$ coefficients are determined to much 
better accuracy than the static dielectric function.

For the other atoms under investigation,
the $C_4$ coefficient read as follows.
For H, Ne, Ar, Kr, and Xe, the results
for $C_4$, in atomic units, read as
$49.99$, $29.52$, $122.9$, $182.8$ and
$303.0$, respectively (at room temperature).

%
%
\section{Conclusions}
\label{sec:conclu}

We have found a unified description of the 
temperature-dependent and frequency-dependent
dielectric function $\epsilon(T_\Delta, \omega)$
of intrinsic (monocrystalline)
silicon, using the LD and CM functional forms, 
augmented by radiation-reaction terms, with
only two generalized oscillator terms
entering the master function given in Eq.~\eqref{master}.
For intrinsic silicon,
we find that both the CM
function $[\epsilon(\omega)-1]/[\epsilon(\omega)+2]$
as well as the LD function
$[\epsilon(\omega) - 1]$ itself can be 
fitted very well to experimental data.
This conclusion is fully consistent with the 
observations made in Ref.~[\onlinecite{OuCa2003}],
where a model without radiation reaction was considered.
The $R^2$ values are greater than $0.99$ for 
either fit, for all temperatures studied here,
as evident from 
Fig.~2 of Ref.~[\onlinecite{MoEtAl2022suppl}].
The CM fit is able to represent 
experimental data marginally better than the LD fit,
consistent with its ability to model the 
local-field effect. The temperature-dependence
of the coefficients of our model is well described 
by simple quadratic forms.

Our fitting, as described in Secs.~\ref{sec2A}
and~\ref{sec2B}, is successful,
and leads to the temperature-dependent
parameters listed in Tables~\ref{table3}
and~\ref{table6}.
These lead to a satisfactory representation
of the dielectric function of intrinsic silicon
in the temperature range 
$0 < T_\Delta < 2.83$,
i.e., $293\,{\rm K} < T < 1123\,{\rm K}$.
The entire problem is of considerable 
interest, and the investigation of a
uniform representation of the dielectric 
function over a wide temperature range
requires a careful evaluation of available
experimental data (see Appendix~\ref{appa}).
The fact that a unified model with analytic
coefficients is able to describe the 
temperature-dependent, and frequency-dependent
dielectric function of intrinsic silicon
over wide ranges of the parameters,
could be interpreted as supporting the
self-consistency of the experimental data 
for the dielectric function, obtained
by various different groups over the 
past two decades~\cite{Je1992,VuEtAl1993,SiHoHu1998,Gr2008,ScEtAl2015}.

We employ the results of our fitting 
in the temperature-dependent evaluation
of the short-range, and long-range,
asymptotics of the atom-surface interaction
potential for helium atoms interacting with 
intrinsic silicon (see Sec.~\ref{sec3}).
We find that the $C_3$ and $C_4$ coefficients
given in Eqs.~\eqref{C3} and~\eqref{C4}
exhibit a moderate temperature dependence
depicted in Figs.~\ref{fig5} and~\ref{fig6}.
Our approach allows us to determine temperature-dependence
$C_3$ and $C_4$ coefficients with a relative accuracy which 
we would like to conservatively estimate as 5\%,
due to the intrinsic uncertainty in the experimental
data, even if the relative difference of the 
$C_3$ and $C_4$ coefficients given in 
Tables~\ref{table7} and~\ref{table8} is smaller than 5\%.
This estimate is supported 
by an error propagation calculation
based on computer algebra~\cite{Wo1999},
which propagates the uncertainty estimate for the 
fit parameters given in Ref.~[\onlinecite{MoEtAl2022suppl}] to the determination
of the $C_3$ and $C_4$ coefficients.
Interestingly, our calculations
imply the existence of a manifest temperature 
dependence of atom-surface interactions,
which goes beyond the 
``thermal discretization'' of the frequencies
of the virtual photons that mediate the atom-surface
interaction, in terms of the Matsubara
frequencies~\cite{DzLiPi1959jetp,DzLiPi1961spu,DzLiPi1961advphys}.

%
%
\section*{Acknowledgments}

Helpful conversations with M. DeKieviet are gratefully acknowledged. C.M., T.D.
and U.D.J. were supported by Grant No. PHY-2110294 from the National Science
Foundation (NSF).  C.A.U. acknowledges support from Grant No. DMR-1810922 from
the National Science Foundation (NSF).

\appendix

%
%
\section{Intricacies of the Dielectric Function}
\label{appA}

%
%
\subsection{Brief Review of Published Data}
\label{appa}

A very brief review of the available experimental 
data for the dielectric function of intrinsic silicon 
might be in order. The subject is interesting because
of a significant dependence on the sample preparation,
with tiny surface impurities 
having a potentially detrimental effect on 
the accuracy of the obtained data.
In view of apparent discrepancies 
among some published data, which will be 
discussed in the following, we here
prefer to use rather recent
compilations of optical properties of silicon;
it is hoped that potential issues with
previous measurements may have been addressed
in the more recent compilations. Specifically,
our sources for experimental data of silicon at
room temperature are Refs.~[\onlinecite{Je1992,Gr2008,ScEtAl2015}].
Our main sources of temperature-dependent data
are Refs.~[\onlinecite{VuEtAl1993,SiHoHu1998}].
We use Ref.~[\onlinecite{SiHoHu1998}]
as a source for the experimental measurements of silicon
at $298$\,K, $523$\,K, $773$\,K, $973$\,K,
and $1123$\,K. For completeness,
Ref.~[\onlinecite{VuEtAl1993}] is used
as a source for experimental measurements of silicon at
$293$\,K (additional data
for room temperature), $373$\,K, $423$\,K, $473$\,K,
$523$\,K, $573$\,K, $623$\,K, $723$\,K.
As a side remark, we can add that the
room temperature data ($293$\,K, from
Refs.~[\onlinecite{Je1992,VuEtAl1993,Gr2008,ScEtAl2015}])
differ only very slightly from the data obtained
for $298$\,K in Ref.~[\onlinecite{SiHoHu1998}].
The data for $298$\,K cover a smaller
frequency range as compared to the
data for $293$\,K; the discussion of the data
available for $298$\,K is relegated to 
Ref.~[\onlinecite{MoEtAl2022suppl}].
We also mention Ref.~[\onlinecite{Gr2008}] for a
discussion of the temperature dependence
of the dielectric function
of silicon, where the linear term of the
coefficients (as a function of the
temperature $T$) is taken into account.

Now, for completeness, let us briefly discuss some apparent 
discrepancies among other data sets.
For example, in Ref.~[\onlinecite{GrKr1995}],
it is pointed out that
``recent measurements \cite{Je1992,KeGr1995},
at both the ultraviolet and infrared
ends of the spectrum have considerably improved the accuracy
of silicon optical data at these wavelengths,
rendering past tabulations~\cite{Ed1985,Gr1987tt,As1988}
and assessments~\cite{BuBrWa1994} largely obsolete.''
Along the same direction, in Sec.~IV.b of Ref.~[\onlinecite{HeEtAl1998}],
it is pointed out that the compilation of
silicon (Si)
optical data in Ref.~[\onlinecite{Ed1985}] relies on two sets of silicon absorption 
values based on the intensity transmission 
measurements originally reported in Refs.~[\onlinecite{Hu1975,MFRo1955}].
Yet, it is pointed out in Sec.~IV.b of Ref.~[\onlinecite{HeEtAl1998}],
that the values reported in Ref.~[\onlinecite{Hu1975}]
spanning 1.2-2\,eV were
obtained from very thin epitaxial films on sapphire and there
is a mismatch by a factor of 5 at 1.28\,eV as compared to
the values reported in Ref.~[\onlinecite{MFRo1955}].

In Ref.~[\onlinecite{HeEtAl1998}], near the start in the Introduction,
it is pointed out that:
``However, even though silicon is one of the
most heavily studied and well-understood materials, the
accuracy of reported optical constant spectra for crystalline 
silicon is still an issue. The original spectroscopic ellipsometry
results for silicon obtained by Aspnes~\cite{AsSt1983} have been questioned
(especially for energies less than 3.4\,eV by the work of
Jellison~\cite{Je1992} using a two-channel polarization modulation
ellipsometer).''
Furthermore, in Ref.~[\onlinecite{HeEtAl1998}], near the start in the Introduction,
it is also pointed out that
the measurements reported in Ref.~[\onlinecite{AsSt1983}] were complicated both
by the difficulty of stripping residual oxide without
roughening the sample and by acquisition of ellipsometric data at an
angle of incidence which pushed the measured ellipsometric
values at smaller photon energies into a sub-optimal region
for the rotating-analyzer ellipsometer (RAE) used. 
It is also pointed out that in 
Ref.~[\onlinecite{Je1992}], a careful 
oxide layer removal procedure was 
profiting from a separate intensity transmission measurement 
used in order to establish the
overlayer thickness.

Finally, we should also mention 
that we have made several unsuccessful attempts to fit the 
data given in Ref.~[\onlinecite{Ed1985}] over the frequency range 
$0 < \hbar \omega < 5 \, {\rm eV}$ with functional 
forms that fulfill the Kramers-Kronig relations.
An inspection reveals that data for 
the imaginary part of the dielectric function
given in Ref.~[\onlinecite{LaEtAl1987}],
and Refs.~[\onlinecite{VuEtAl1993,SiHoHu1998}], exceeds the
values given in Ref.~[\onlinecite{Ed1985}] in the frequency range
$1 < \hbar \omega < 2 \, {\rm eV}$ 
by almost a factor two.
The newer data given in Refs.~[\onlinecite{VuEtAl1993,SiHoHu1998}] 
is amenable to a fit using a consistent functional form, 
as detailed in this current study.
Furthermore, we note that no actual data pairs 
of frequency and real and imaginary part of the 
dielectric function are given in Ref.~[\onlinecite{LaEtAl1987}].
However, a quantitative inspection of the curves given Figs.~2, 3, and 4 
of Ref.~[\onlinecite{LaEtAl1987}] leads to the 
conclusion that the data on which the
Ref.~[\onlinecite{LaEtAl1987}] is based, are in agreement 
with the analysis presented in the current investigation.

The availability of convenient, consistent, simple functional
forms to describe the frequency-dependent, 
and temperature-dependent dielectric function
of intrinsic silicon, as derived here, should thus be 
of considerable interest to the community.

%
%
\subsection{Lorentz-Dirac Model}
\label{appb}

Let us start from Eq.~(2) Ref.~[\onlinecite{PrKa2014}],
which describes the acceleration $\vec a$ on a charge carrier particle 
of charge $q$ in terms of the Lorentz-Dirac formalism,
\begin{equation}
\vec a = \frac{1}{m} \vec F_{\rm ext} + t_0 \, \dot{\vec a} \,.
\end{equation}
The latter term describes radiation reaction.
A discussion of the Lorentz-Dirac equation can be 
found in Sec.~8.6.2 of Ref.~[\onlinecite{Je2017book}].
The radiation reaction time is 
[see Eq.~(3) of Ref.~[\onlinecite{PrKa2014}]]
\begin{equation}
t_0 = \frac{q^2}{6 \pi \epsilon_0 m c^3} \,.
\end{equation}
One defines a characteristic acceleration 
$\vec a_c$ and a characteristic time 
scale $t_c$ through the formulas
[see Eqs.~(4) and~(5) of Ref.~[\onlinecite{PrKa2014}]]
\begin{equation}
\vec a_c = \frac{1}{m} \, \vec F_{\rm ext} \,,
\qquad
\dot{\vec a}_c = \frac{\vec a_c}{t_c} \,,
\end{equation}
where $\vec F_{\rm ext}$ is the external force.
Then, according to Eq.~(6) of Ref.~[\onlinecite{PrKa2014}],
one defines 
\begin{equation}
t_{\rm light} = s_q/c
\end{equation}
as the time it takes light to travel a characteristic 
distance $s_q$ which 
could be chosen as the size of the charge distribution,
or, from a classical point of view, 
as the classical electron radius obtained 
by equating the electron rest mass 
with the electrostatic self-energy of the 
electron's charge distribution, taken as centered
on a sphere of radius $s_q$.
Then, according to Eq.~(7) of Ref.~[\onlinecite{PrKa2014}],
one has
\begin{equation}
m \sim m_{\rm em} \sim \frac{q^2}{4 \pi \epsilon_0 s_q} \,,
\end{equation}
where $m_{\rm em}$ is the self-energy (self-mass) 
of the electron.

One assumes a not-too-fast change in the acceleration,
i.e., a not-too-abrupt dynamical change,
\begin{equation}
t_c \gg t_{\rm light} \,.
\end{equation}
Under the observation
[see Eq.~(9) of Ref.~[\onlinecite{PrKa2014}]]
that $t_0 \sim t_{\rm light}$,
one derives the condition
[see Eq.~(10) of Ref.~[\onlinecite{PrKa2014}]]
\begin{equation}
t_0 \sim t_{\rm light} \ll t_c \,,
\qquad
t_0/t_c \ll 1 \,,
\end{equation}
under which the authors of Ref.~[\onlinecite{PrKa2014}] 
arrive at the following formula
[see Eq.~(25) of Ref.~[\onlinecite{PrKa2014}]]
for the polarization density $\vec P$
in the sample,
\begin{multline}
\label{poldens}
\frac{\dd^2 \vec P}{\dd t^2} 
+ \left( 2 \Gamma + t_0 \, \omega_r^2 
- \frac{q^2 t_0 N}{3 m \epsilon_0} 
\right)\, 
\frac{\dd \vec P}{\dd t} 
\\
+ \left( \omega_r^2 
- \frac{q^2 N}{3 m \epsilon_0} \right)  \vec P
= \frac{q^2 N}{m} \, \vec E + 
\frac{q^2 N t_0}{m} \, \frac{\dd \vec E}{\dd t} \,.
\end{multline}
From Ref.~[\onlinecite{PrKa2014}], one can see that 
$\Gamma$ is the damping rate associated with the 
frictional force between atoms, $\omega_r$ is the natural 
frequency of the restoring force, $N$ is the number of 
atoms per unit volume, and $q$ is the charge of the 
electron.
The transformation to Fourier space proceeds 
by writing
\begin{equation}
\vec P(t) = \int \frac{\dd \omega}{2 \pi} \,
\ee^{-\ii \omega t} \, \vec P(\omega) \,,
\end{equation}
so that, in Fourier space, one replaces
$\dd/\dd t \to -\ii \omega$. 

Setting $\vec P(\omega) = \epsilon(\omega) \, 
\epsilon_0 \, \vec E(\omega)$, 
one then arrives at the following formula,
\begin{equation}
\label{res2}
\epsilon(\omega) = \epsilon_\infty +
\frac{a_0 - \ii \omega \, a_1}
{b_0 - \omega^2 - \ii \omega \, b_1}\,.
\end{equation}
With reference to Eq.~\eqref{poldens} and
Eq.~(27) from Ref.~[\onlinecite{PrKa2014}],
the parameters are identified as follows,
\begin{subequations}
\begin{align}
b_1 =& \; 2\Gamma+t_0 \, \omega_r^2-\frac{q^2t_0N}{3m\epsilon_0}\,,
& b_0 & =\omega_r^2-\frac{q^2N}{3m\epsilon_0} \,,
\\
a_0=& \frac{q^2N}{m}\,,
& a_1 & =\frac{q^2t_0N}{m} \,.
\end{align}
\end{subequations}

In Eq.~\eqref{res2}, the signs of the terms multiplying 
$a_1$ and $b_1$ are inverted as compared to 
Eq.~(28) of Ref.~[\onlinecite{PrKa2014}],
presumably due to a typographical error in 
Ref.~[\onlinecite{PrKa2014}].
Note that, upon using the functional
form~\eqref{res2}, $a_1$ and $b_1$ are obtained as positive
rather than negative quantities in our fitting procedure,
for silicon, supporting the functional form 
indicated in Eq.~\eqref{res2} (with positive
terms $a_1$ and $b_1$). 
This finding also is in line with the functional form 
used in Ref.~[\onlinecite{DeJo2012}].

\begin{table}[t]
\caption{\label{table9} Energy differences 
are given $E_{n0}=E_n-E_0$ 
between the reference $1S$ ground state of hydrogen (H) and excited 
states (${\rm nP}$). The fine-structure and Lamb shift are 
not resolved. We also list corresponding oscillator strengths
$f_{n0}$ for the first $10$ excited states. The data includes 
the reduced-mass correction (the oscillator strength 
scales with the first power of $\mu/m_e$, where $\mu$ is the 
reduced mass of hydrogen, and $m_e$ is the electron mass.)
All entries are in 
agreement with the data compilation given in Ref.~[\onlinecite{WiFu2009}].}
\renewcommand{\arraystretch}{1.25}
\begin{center}
\begin{minipage}{0.95\linewidth}
\begin{tabular}
{c@{\hspace{.7cm}}c@{\hspace{.7cm}}c}
\hline
\hline
$n$ & $E_{n0}\,[E_h]$ & $f_{n0}\,[e^2 a_0^2 E_h ]$ \\ 
\hline
2  & 0.37480 & 0.41640 \\
3  & 0.44421 & 0.07914 \\
4  & 0.46850 & 0.02901 \\
5  & 0.47974 & 0.01395 \\
6  & 0.48585 & 0.00780 \\
7  & 0.48954 & 0.00482 \\
8  & 0.49193 & 0.00319 \\
9  & 0.49356 & 0.00222 \\
10 & 0.49473 & 0.00161 \\
11 & 0.49560 & 0.00120 \\
\hline
\hline
\end{tabular}
\end{minipage}
\end{center}
\end{table}

\begin{figure}[t]
\begin{center}
\includegraphics[width=.9\linewidth]{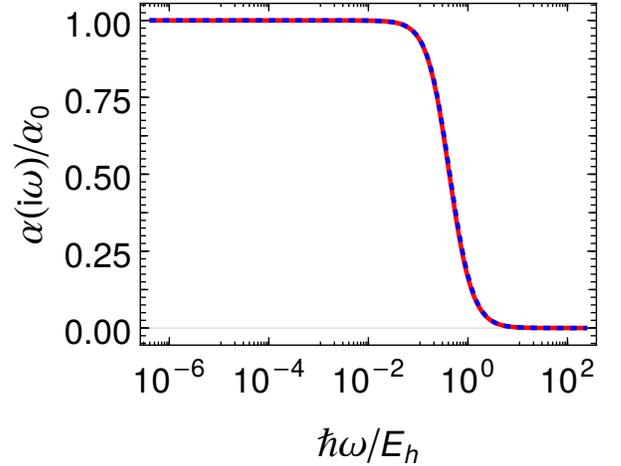}
\end{center}
\caption{\label{fig7}
The dynamic polarizability of atomic hydrogen is plotted 
as a function of the imaginary driving frequency.
The exact solution given in Eq.~\eqref{exactpol} (red curve) 
is compared with the approximation given in Eq.~\eqref{modelpol}
(blue-dotted curve),
which is based on the oscillator strengths listed in
Table~\ref{table9}. The static polarizability of 
hydrogen is $\alpha_0=9/2 e^2 a_0^2/E_h$,
which is $9/2$ in atomic units. 
The relative difference between the exact 
values and the approximation can be found in Fig.~\ref{fig8}.}
\end{figure}

\begin{figure}[t]
\begin{center}
\includegraphics[width=.9\columnwidth]{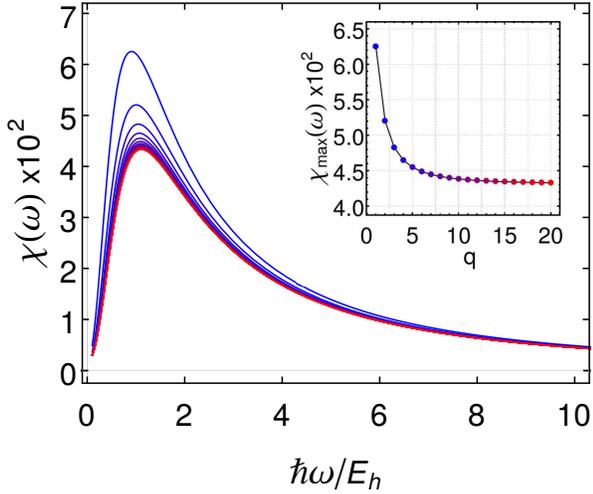}
\end{center}
\caption{\label{fig8}
The relative difference 
$\chi(\omega)$, described in Eq.~\eqref{eq:relDif},
between the exact expression for the dynamic
polarizability of hydrogen 
given in Eq.~\eqref{exactpol} and the discrete model 
given in Eq.~\eqref{modelpol},
is plotted as a
function of the driving frequency.
Here $q$ is the number of discrete oscillator strengths
included in the discrete model,
as given in Table~\ref{table9}.}
\end{figure}

\begin{figure}[t]
\begin{center}
\includegraphics[width=.9\columnwidth]{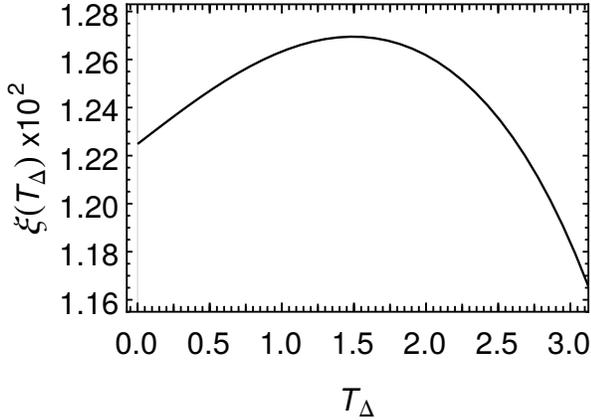}
\end{center}
\caption{\label{fig9}
The relative difference $\xi(T_\Delta)$,
defined in Eq.~\eqref{defxi},
between the short-distance Casimir-Polder parameter 
$C_3$ evaluated using the 
exact expression for the dynamic
polarizability given in Eq.~\eqref{exactpol} and the discrete model 
given in Eq.~\eqref{modelpol}
is plotted as a function of temperature for hydrogen interacting 
with silicon. Note that the plot range 
on the ordinate axis is restricted.}
\end{figure}
%
%
\section{Dynamic Polarizability}
\label{appB}

%
%
\subsection{General Algorithm}
\label{sec:genAlg}

We aim to delineate a rather general  
algorithm here which allows one to 
calculate the dynamic polarizability of an
atom at imaginary driving frequency,
$\alpha(\ii \omega)$, 
based on the knowledge of 
the oscillator strengths of a 
few low-lying transitions,
and additional input from the 
known asymptotic behavior of the 
polarizability for large driving frequency,
to be derived from sum rules.
The algorithm should be 
accurate to a few percent over the 
entire frequency range $0 < \omega < \infty$
and thus sufficient for the 
calculation of atom-surface interactions,
where the dominant source of uncertainty 
comes from the dielectric function (see Sec.~\ref{sec3}).

The approach is to first collect, 
from databases~\cite{NISTSPECTRA2014},
the transition energies and oscillator strengths
of a few low-lying transitions.
This collection immediately 
allows to describe the 
frequency dependence of the dynamic polarizability 
for low excitation frequency argument.
In order to model the contribution of the 
continuum states of the atom, 
we add one more virtual transition to a 
``pseudo-level'', which is energetically 
positioned in the continuum.
The oscillator strength is matched
against the Thomas-Reiche-Kuhn (TRK) sum 
rule~\cite{ReTh1925zahl,Ku1925} sum rule, 
and the energy of the pseudo-level is 
adjusted so that the correct
overall low-frequency (``static'') limit 
of the polarizability is recovered.
Because we only consider imaginary frequencies, 
we are far enough away from any atomic resonance 
that we do not need to worry about 
the decay width of the state, i.e., about 
the imaginary part of the energy that otherwise 
enters the polarizability.

The atomic polarizability is defined as
(see Refs.~[\onlinecite{YaBaDaDr1996,LaDKJe2010pra}])
\begin{equation}
\label{eq:atomicPol}
\alpha\left(\omega\right)=
\sum_n^\infty
\frac{e^2 \, a_0^2 \, E_h \, f_{n0}}{E_{n0}^2-\left(\hbar\omega\right)^2} \,,
\end{equation}
where $f_{n0}$ is the oscillator strength of the atom, 
measured in atomic units, and 
$E_{n0}\equiv E_n -E_0$ is the energy difference 
between the virtual and exited states $\ket{\psi_n}$.
We note that the oscillator strength is 
used, in atomic physics, as a dimensionless
quantity (for an excellent overview of pertinent 
conventions, see Ref.~[\onlinecite{Hi1982}].)
The sum is carried out over all of the 
discrete states as well as the continuous spectrum.

In order to approximate the atomic polarizability
with our model polarizability, $\alpha(\omega) \approx 
\alpha_m(\omega)$, we divide the infinite
sum in Eq.~(\ref{eq:atomicPol}) into two parts,
\begin{equation}
\label{modelpol}
\alpha(\omega) \approx 
\alpha_m(\omega) = \alpha_d(\omega) + \alpha_{c}(\omega) \,,
\end{equation}
in which $\alpha_d(\omega)$ is the sum over 
the terms from the first $q$ discrete (bound) states
\begin{equation}
\label{eq:polApprx}
\alpha_d(\omega) = \sum_{n=1}^q 
\frac{e^2 a_0^2 E_h \, f_{n0}}{E_{n0}^2-\left(\hbar\omega\right)^2} \,,
\end{equation}
where we neglect the width of the virtual states,
anticipating that our final aim will be 
to evaluate the polarizability at imaginary driving 
frequency (where the decay width terms are 
negligible for our purposes). 

Let us denote by $\alpha_c$ the contribution of the 
continuum states. We model the contribution 
$\alpha_c(\omega)$, using a pseudo-level,
as follows,
\begin{equation}
\label{eq:polCont}
\alpha_{\mathrm{c}}(\omega) =
\frac{e^2 a_0^2 E_h \, f_\infty }{E_{\infty}^2-\left(\hbar\omega\right)^2} \,.
\end{equation}
The oscillator strength of the additional ``continuum'' 
level $f_\infty$ is found by requiring that sum over all 
oscillator strengths obey the TRK sum rule 
which in SI mksA units can be expressed as 
\begin{equation}
\sum_n f_{n0}=N \,,
\end{equation}
where $N$ is the total number of electrons in the system,
and $n$ runs over all virtual levels
(discrete and continuum). Therefore,
the matching condition for the oscillator strength 
of the continuum pseudo-level is 
\begin{equation}
f_\infty\approx N-\sum_{n=1}^q f_{n0} \,.
\end{equation}
The energy position of the additional ``continuum'' level $E_\infty$ can be
found by requiring our {\em ansatz} to reproduce known numerical
values of the static polarizability,
\begin{equation}
E_\infty^2=\frac{f_\infty}{\alpha(0)- \alpha_{\mathrm{q}}(0)} \,.
\end{equation}
This algorithm will be applied to hydrogen,
before being generalized to other atoms.

%
%
\subsection{Hydrogen}

Because the dynamic polarizability of hydrogen can be 
calculated analytically~\cite{LeLe1996,Ya2003,Pa1993,JePa1996}, 
a comparison of the complete result to that found using the 
algorithm described above can be used as a measure of the validity of
our {\em ansatz}. We start from the analytic solution 
for the dielectric function for the ground state of hydrogen
as a function of $\omega$ as described 
in Refs.~[\onlinecite{Pa1993,JePa1996,LeLe1996}],
\begin{equation}
\label{exactpol}
\alpha(\omega)=
\frac{e^2 \, \hbar^2}{\alpha^4 \, \mu^3 \, c^4}
\left[ Q(\omega) + Q(-\omega) \right] \,,
\end{equation}
where $\mu$ is the reduced mass of hydrogen,
$\alpha$ is the fine-structure constant,
$c$ is the speed of light, $e$ is the elementary charge,
and $\hbar$ is Planck's unit of action.
The matrix element $Q = Q(\omega)$ is given as follows,
\begin{equation}
Q(\omega) = \frac{E_h}{e^2 \, a_0^2} \,
\left< {\rm 1S} \left| \vec r \frac{1}{H_S - E_{\rm 1S} + \omega} \vec r \right|
{\rm 1S} \right> \,,
\end{equation}
where ${\rm 1S}$ denotes the ground state of hydrogen,
the scalar product is understood for 
the position operators $\vec r$, 
$H_S$ is the Schr\"{o}dinger-Coulomb Hamiltonian, and
$E_{\rm 1S}$ is the ground-state energy,

The $Q$ matrix elements are dimensionless
and can be expressed in terms of the 
dimensionless photon energy variable
\begin{equation}
t = t(\omega) = 
\left(1 + \frac{2 \hbar\omega}{E_h}\right)^{-1/2} \,,
\end{equation}
and
\begin{multline}
Q(\omega) = \frac{2t^2}{3(1-t)^5(1+t)^4}
\, \left[ 3-3t-12t^2+12t^3
\right.
\\
\left.  +19t^4 - 19t^5-26t^6-38t^7 \right]
\\
+ \frac{256t^9}{3(1+t)^5(1-t)^5}
\,{}_2F_1\left(1,-t,1-t,\left(\frac{1-t}{1+t}\right)^2\right) \,,
\end{multline}
and it is understood that $t \equiv t(\omega)$.
Here, ${}_2 F_1$ is the Gaussian hypergeometric function.
In Table~\ref{table9}, we collect oscillator 
strengths for the first ten dipole-allowed 
hydrogen transitions from the reference ground state
to excited ${\rm nP}$ states, with $n=2,\dots,11$. 
One can verify that the oscillator strengths, for general $n$,
obey the following general formula,
\begin{equation}
\label{eq:oscillatorstrengthhydrogen}
f_{n0} = \frac{256 n^5}{3 (n^2-1)^4} \left( \frac{n-1}{n+1} \right)^{2n}
\end{equation}
which can be derived starting from Eq.~(6.133) of 
Ref.~[\onlinecite{Je2017book}]. The
angular integral in that expression can be calculated directly while
Eq.~(7.414.7) of Ref.~[\onlinecite{GrRy1994}]
can be used to evaluate the radial
integral. The Gaussian hypergeometric function that appears in the result can
be expressed in closed form. The result given in 
Eq.~\eqref{eq:oscillatorstrengthhydrogen},
upon the inclusion of reduced-mass effects, reproduces all
data collected in Table~\ref{table9}, originally collected 
from Ref.~[\onlinecite{WiFu2009}].

The panel in
Fig.~\ref{fig7} shows the numerical results for the dynamic
polarizability as a function of $\omega$. Numerical results from the proposed
algorithm for the first ten energy differences and oscillator strengths $f_{n0}$
collected in Table.~\ref{table9} (blue-dotted line), are nearly identical
to those from the analytic solution in Eq.~\eqref{exactpol} (red-dotted
line). A closer look, as described by
Fig.~\ref{fig8}, reveals a peak in the relative difference $\chi(\omega)$ of 
the exact dynamic polarizability of hydrogen 
given in Eq.~\eqref{exactpol}, and the model polarizability given in 
Eq.~\eqref{modelpol}, 
\begin{equation}
\label{eq:relDif}
\chi(\omega) = \frac{\alpha_m(\ii \omega) - 
\alpha(\ii \omega)}{\alpha(\ii \omega)} \,,
\end{equation}
at a driving frequency of about one atomic unit.
However, the peak relative difference occurs in a region
where the absolute value of the polarizability
has already dropped to about one tenth of 
its static value (see Fig.~\ref{fig7}), and 
is thus less than $1$\,\% when divided by the static 
polarizability. 
Note that the plot pertains to imaginary driving frequencies,
so that the bound-state poles remain invisible.

Let us compare, for the Lorentz-Dirac fit (at room
temperature), the result for $C_3$, for hydrogen
interacting with silicon, evaluated in terms
of the model polarizability~\eqref{modelpol},
to the result obtained using the exact polarizability,
given in Eq.~\eqref{exactpol}. We define
\begin{equation}
C^{(m)}_3(T_\Delta) = \frac{\hbar}{16\pi^2\epsilon_0}
\int_0^\infty \dd\omega \; \alpha_m(\ii\omega) \;
\frac{\epsilon_\LD(T_\Delta,\ii\omega)-1}%
{\epsilon_\LD(T_\Delta,\ii\omega)+1} \,,
\end{equation}
as the result obtained from the model polarizability,
and 
\begin{equation}
C^{(e)}_3(T_\Delta) = \frac{\hbar}{16\pi^2\epsilon_0}
\int_0^\infty \dd\omega \; \alpha(\ii\omega) \;
\frac{\epsilon_\LD(T_\Delta,\ii\omega)-1}%
{\epsilon_\LD(T_\Delta,\ii\omega)+1} \,,
\end{equation}
as the result obtained using the 
exact polarizability.  Then, the relative difference
is
\begin{equation}
\label{defxi}
\xi(T_\Delta) = \frac{ C^{(m)}_3(T_\Delta) -
C^{(e)}_3(T_\Delta) }{ C^{(e)}_3(T_\Delta) } \,,
\end{equation}
and it is plotted in Fig.~\ref{fig9}.
The difference of about 1\,\% is negligible 
on the level of the uncertainty in the determination
of $C_3$ implied by the 
dielectric function.

%
%
\subsection{Other Elements}

Just as for hydrogen, 
the atomic polarizability of helium may
be calculated according to the algorithm
outlined in Sec.~\ref{sec:genAlg}.
The static polarizability of helium is
$\alpha(\omega = 0) = 1.383\, e^2 \, a_0^2/E_h$ \cite{YaBaDaDr1996,PaSa2000},
which is equivalent to a numerical value of $1.383$
in atomic units.
For helium, extensive calculations are available
(see Refs.~[\onlinecite{KaSa2012,MaSt2003,GlWe1976,ChKw1968,ErRe1985,BiLa1988}]).
Data for other elements can easily be found in the
NIST database, which is available online (see Ref.~[\onlinecite{NISTSPECTRA2014}]).

Furthermore, additional data on oscillator strengths 
is available for other atoms of interest,
from Refs.~[\onlinecite{SaSa1977,WiBrDaHeKo1989,Se1998,Mo2003atm,Kr2013}].

\end{document}